\documentclass[final,5p,times]{elsarticle}

\usepackage[draft]{hyperref}
\usepackage{lineno}
\usepackage{etoolbox}

\modulolinenumbers[2]
\journal{Journal of \LaTeX\ Templates}

\usepackage{comment}                        %
\usepackage{colortbl}                       %
\usepackage{aas_macros}                     %
\usepackage{alphalph}                       %
\usepackage[normalem]{ulem}                 %
\usepackage{threeparttable}                 %
\usepackage{multirow}                       %
\usepackage{float}                          %
\usepackage{lscape}                         %
\usepackage{footnote}                       %
\usepackage{tabularx}                       %
\usepackage{etoolbox}                       %
\usepackage[colorinlistoftodos]{todonotes}  %
\usepackage[printwatermark]{xwatermark}     %
\usepackage{xcolor}                         %
\usepackage{graphicx}                       %
\usepackage{lipsum}                         %

\usepackage{eso-pic}

\AddToShipoutPictureBG*{%
  \AtPageUpperLeft{%
    \hspace{0.75\paperwidth}%
    \raisebox{-3.5\baselineskip}{%
      \makebox[0pt][l]{\textnormal{DES 2018-0355}}
}}}%

\AddToShipoutPictureBG*{%
  \AtPageUpperLeft{%
    \hspace{0.75\paperwidth}%
    \raisebox{-4.5\baselineskip}{%
      \makebox[0pt][l]{\textnormal{FERMILAB-PUB-17-043}}
}}}%

\definecolor{light-gray}{gray}{0.3}
\newcommand{\comp}[1]{\texttt{#1}} 

\biboptions{authoryear}

\begin{document}
\begin{frontmatter}


%
\title{DES Science Portal: Computing Photometric Redshifts}
%


\author[on,linea]{J.~Gschwend\corref{cor1} } 
\ead{julia@linea.gov.br}
\author[on,linea]{A.~Carnero~Rossel}
\author[on,linea]{R.~L.~C.~Ogando}
\author[lsst,linea]{A.~Fausti~Neto}
\author[on,linea]{M.~A.~G.~Maia}
\author[on,linea]{L.~N.~da~Costa}
\author[ifusp,linea]{M.~Lima}
\author[on,linea]{P.~Pellegrini}
\author[linea,cefet]{R.~Campisano}
\author[on,linea]{C.~Singulani}
\author[linea]{C.~Adean}
\author[lagrange,linea]{C.~Benoist}
\author[linea]{M.~Aguena}
\author[uillinois,ncsa]{M.~Carrasco~Kind} %
\author[queensland,caastro]{T.~M.~Davis} %
\author[ciemat]{J.~de~Vicente}%
\author[ucl]{W.~G.~Hartley}%
\author[munich]{B.~Hoyle}%
\author[ucl,fermilab]{A.~Palmese}%

\author[desy]{I.~Sadeh}%


\author[ctio]{T.~M.~C.~Abbott}
\author[ucl,rhodes]{F.~B.~Abdalla}%
\author[fermilab]{S.~Allam}%
\author[fermilab]{J.~Annis}%
\author[queensland,caastro,swinburne]{J.~Asorey}%

\author[ucl]{D.~Brooks}%

\author[queensland]{J.~Calcino}%
\author[inaf]{D.~Carollo}%
\author[ieec,ice-csic]{F.~J.~Castander}%

\author[upenn]{C.~B.~D'Andrea}%
\author[hyderabad]{S.~Desai}%

\author[michast,michphy]{A.~E.~Evrard}%

\author[ieec,ice-csic]{P.~Fosalba}%
\author[fermilab,kav-chicago]{J.~Frieman}%

\author[uam]{J.~Garc\'ia-Bellido}%
\author[swinburne,caastro]{K.~Glazebrook}%
\author[michast,michphy]{D.~W.~Gerdes}%
\author[uillinois,ncsa]{R.~A.~Gruendl}%
\author[fermilab]{G.~Gutierrez}%

\author[queensland,caastro]{S.~Hinton}%
\author[santacruz]{D.~L.~Hollowood}%
\author[osu-ccapp,osu-ast]{K.~Honscheid}
\author[queensland]{J.~K.~Hoormann}%

\author[harv]{D.~J.~James}%

\author[aao]{K.~Kuehn}%
\author[fermilab]{N.~Kuropatkin}%

\author[ucl]{O.~Lahav}%
\author[usydney]{G.~Lewis}%
\author[aao]{C.~Lidman}%
\author[fermilab]{H.~Lin}%

\author[port]{E.~Macaulay}%
\author[tam]{J.~Marshall}%
\author[princeton]{P.~Melchior}%
\author[icrea,ifae]{R.~Miquel}%
\author[aus-nat-uni,caastro]{A.~M\"oller}%

\author[jpl,asp]{A.~A.~Plazas}%


\author[ciemat]{E.~Sanchez}%
\author[ufrgs,linea]{B.~Santiago}%
\author[fermilab]{V.~Scarpine}%
\author[slac]{R.~H.~Schindler}%
\author[ciemat]{I.~Sevilla-Noarbe}%
\author[southampton]{M.~Smith}%
\author[unicamp,linea]{F.~Sobreira}%
\author[aus-nat-uni,caastro]{N.~E.~Sommer}%
\author[ornl]{E.~Suchyta}%
\author[ncsa]{M.~E.~C.~Swanson}%

\author[michphy]{G.~Tarle}%
\author[aus-nat-uni,caastro]{B.~E.~Tucker}%
\author[fermilab]{D.~L.~Tucker}%

\author[purple]{S.~Uddin}%

\author[ctio]{A.~R.~Walker}%

%

\cortext[cor1]{Corresponding author}
\address[on] {Observat\'orio Nacional, Rua General Jos\'e Cristino, 77, Rio de Janeiro, RJ, 20921-400, Brazil} 
\address[linea] {Laborat\'orio Interinstitucional de e-Astronomia - LIneA, Rua General Jos\'e Cristino, 77, Rio de Janeiro, RJ, 20921-400, Brazil}
\address[lsst]{LSST Project Management Office, Tucson, AZ, USA}
\address[ifusp] {Departamento de F\'isica Matem\'atica, Instituto de F\'isica, Universidade de S\~ao Paulo, CP 66318, S\~ao Paulo, SP, 05314-970, Brazil }
\address[cefet] {Centro Federal de Educa\c c\~ ao Tecnol\'ogica Celso Suckow da Fonseca - CEFET/RJ, Av. Maracan\~a, 229, Rio de Janeiro, RJ, 20271-110, Brazil}
\address[lagrange]{Laboratoire Lagrange, Universit\'e C\^ote d’Azur, Observatoire de la C\^ote d’Azur, CNRS, Blvd de l’Observatoire, CS 34229, 06304 Nice cedex 4, France}
%
\address[uillinois] {Department of Astronomy, University of Illinois, 1002 W. Green Street, Urbana, IL 61801, USA}
\address[ncsa] {National Center for Supercomputing Applications, 1205 West Clark St., Urbana, IL 61801, USA}
\address[queensland] {School of Mathematics and Physics, University of Queensland, QLD 4072, Australia}
\address[caastro] {ARC Centre of Excellence for All-sky Astrophysics (CAASTRO), Australia}
\address[ciemat] {Centro de Investigaciones Energ\'eticas, Medioambientales y Tecnol\'ogicas (CIEMAT), Avda. Complutense 40, E-28040, Madrid, Spain}
\address[ucl] {Department of Physics \& Astronomy, University College London, Gower Street, London, WC1E 6BT, UK}
\address[munich] {Universit\"ats-Sternwarte, Fakult\"at f\"ur Physik, Ludwig-Maximilians Universit\"at M\"unchen, Scheinerstr. 1, D-81679 M\"unchen, Germany}
\address[fermilab] {Fermi National Accelerator Laboratory, P. O. Box 500, Batavia, IL 60510, USA}
\address[desy]{Deutsches Elektronen-Synchrotron (DESY),
Platanenallee 6, 15738 Zeuthen, Germany.}
%
%
\address[ctio] {Cerro Tololo Inter-American Observatory, National Optical Astronomy Observatory, Casilla 603, La Serena, Chile}
\address[rhodes] {Department of Physics and Electronics, Rhodes University, PO Box 94, Grahamstown, 6140, South Africa}
\address[swinburne]{Centre for Astrophysics and Supercomputing, Swinburne University of Technology, PO Box 218, Hawthorn, VIC 3122, Australia}
\address[inaf]{INAF - Astrophysical Observatory of Turin, Italy}
\address[ieec]{Institut d'Estudis Espacials de Catalunya (IEEC), 08193 Barcelona, Spain}
\address[ice-csic]{Institute of Space Sciences (ICE, CSIC),  Campus UAB, Carrer de Can Magrans, s/n,  08193 Barcelona, Spain}
\address[upenn] {Department of Physics and Astronomy, University of Pennsylvania, Philadelphia, PA 19104, USA}
\address[hyderabad] {Department of Physics, IIT Hyderabad, Kandi, Telangana 502285, India}
\address[michast] {Department of Astronomy, University of Michigan, Ann Arbor, MI 48109, USA}
\address[michphy] {Department of Physics, University of Michigan, Ann Arbor, MI 48109, USA}
\address[kav-chicago] {Kavli Institute for Cosmological Physics, University of Chicago, Chicago, IL 60637, USA}
\address[uam] {Instituto de Fisica Teorica UAM/CSIC, Universidad Autonoma de Madrid, 28049 Madrid, Spain}
\address[santacruz]{Santa Cruz Institute for Particle Physics, Santa Cruz, CA 95064, USA}
\address[osu-ccapp] {Center for Cosmology and Astro-Particle Physics, The Ohio State University, Columbus, OH 43210, USA}
\address[osu-ast] {Department of Astronomy, The Ohio State University, Columbus, OH 43210, USA}
\address[harv] {Harvard-Smithsonian Center for Astrophysics, Cambridge, MA 02138, USA}
\address[aao] {Australian Astronomical Observatory, North Ryde, NSW 2113, Australia}
\address[usydney] {Sydney Institute for Astronomy, School of Physics, A28, The University of Sydney, NSW 2006, Australia}
\address[port]{Institute of Cosmology \& Gravitation, University of Portsmouth, Portsmouth, PO1 3FX, UK}
\address[tam] {George P. and Cynthia Woods Mitchell Institute for Fundamental Physics and Astronomy, and Department of Physics and Astronomy, Texas A\&M University, College Station, TX 77843, USA}
\address[princeton]{Department of Astrophysical Sciences, Princeton University, Peyton Hall, Princeton, NJ 08544, USA}
\address[icrea] {Instituci\'o Catalana de Recerca i Estudis Avan\c{c}ats, E-08010 Barcelona, Spain}
\address[ifae] {Institut de F\'{\i}sica d'Altes Energies (IFAE), The Barcelona Institute of Science and Technology, Campus UAB, 08193 Bellaterra (Barcelona) Spain}
\address[aus-nat-uni] {Research School of Astronomy and Astrophysics, Australian National University, Canberra, ACT 2611, Australia}
\address[jpl] {Jet Propulsion Laboratory, California Institute of Technology, 4800 Oak Grove Dr., Pasadena, CA 91109, USA}
\address[asp]{Astronomical Society of the Pacific, 100 N Main St., Suite 15, Edwarsville, IL 62025, USA}
\address[ufrgs] {Instituto de F\'\i sica, UFRGS, Caixa Postal 15051, Porto Alegre, RS - 91501-970, Brazil}
\address[slac]{SLAC National Accelerator Laboratory, Menlo Park, CA 94025, USA}
\address[southampton] {School of Physics and Astronomy, University of Southampton, Southampton, SO17 1BJ, UK}
\address[unicamp]{Instituto de F\'isica Gleb Wataghin, Universidade Estadual de Campinas, Campinas, SP, 13083-859, Brazil}
\address[ornl] {Computer Science and Mathematics Division, Oak Ridge National Laboratory, Oak Ridge, TN 37831, USA}
\address[purple] {Purple Mountain Observatory, Chinese Academy of Sciences, Nanjing, Jiangsu, China}


\end{frontmatter}

\section*{Abstract}

A significant challenge facing photometric surveys for cosmological purposes is the need to produce reliable redshift estimates. The estimation of photometric redshifts (photo-$z$s) has been consolidated as the standard strategy to bypass the high production costs and incompleteness of spectroscopic redshift samples.  Training-based photo-$z$ methods require the preparation of a high-quality list of spectroscopic redshifts, which needs to be constantly updated. The photo-$z$ training, validation, and estimation must be performed in a consistent and reproducible way in order to accomplish the scientific requirements. To meet this purpose, we developed an integrated web-based data interface that not only provides the framework to carry out the above steps in a systematic way, enabling the ease testing and comparison of different algorithms, but also addresses the processing requirements by parallelizing the calculation in a transparent way for the user. This framework called the Science Portal (hereafter Portal) was developed in the context the Dark Energy Survey (DES) to facilitate scientific analysis. In this paper, we show how the Portal can provide a reliable environment to access vast data sets, provide validation algorithms and metrics, even in the case of multiple photo-$z$s methods. It is possible to maintain the provenance between the steps of a chain of workflows while ensuring reproducibility of the results. We illustrate how the Portal can be used to provide photo-$z$ estimates using the DES first year (Y1A1) data. While the DES collaboration is still developing techniques to obtain more precise photo-$z$s, having a structured framework like the one presented here is critical for the systematic vetting of DES algorithmic improvements and the consistent production of photo-$z$s in future DES releases.

%
%
\section*{Keywords}

{astronomical databases: catalogs, surveys} --
{methods: data analysis} --
{galaxies: distances and redshifts, statistics} 

%


    \section{Introduction}
    \label{sec:intro}

In the last few decades, large galaxy surveys have become one of the main research tools in astronomy, in particular, for the study of cosmology. The need for increasing statistical samples and depths have encouraged the design and construction of deeper, wider, and more sensitive surveys. These projects are generating vast amounts of data, bringing astronomy into the realm of ``big data'', which increases the challenges associated with cosmological analyses.

One of these projects is the Dark Energy Survey {\citep[DES,][]{Fla05, Abb16b}, a 5--year program to carry out two distinct surveys. The wide--angle survey covers 5,000 deg$^2$ of the southern sky in five ($grizY$) filters to a nominal magnitude limit of $\sim$24 in most bands. Also, there is a deep survey ($i \sim$26) of about 30 deg$^2$ in four filters ($griz$) with a well--defined cadence to search for type--Ia Supernovae (SNe Ia) \citep{Kes15}. The primary goal of DES is to constrain the nature of dark energy through the combination of four observational probes, namely baryon acoustic oscillations, counts of galaxy clusters, weak gravitational lensing, and determination of distances of SNe. Besides, many other fields of astrophysics benefit from the large data set generated by the survey, as detailed by \citet{Abb16b}.

The constraining power of cosmological results provided by DES will strongly depend on the ability to estimate reliable photometric redshifts \citep[photo-$z$, e.g.,][]{Hut04, Ma06, Lim07, Ma08, Hea10, Cun14, Geo14}. In fact, the computation of accurate photo-$z$s has been one of the major concerns of the collaboration, which has spurred the implementation and testing of several algorithms. For instance, \citet{San14} addressed the performance of several codes when applied to the DES science verification data (SVA1), while \citet{Ban15} discussed the effect of using infrared data. More recently, \citet{Bon16} examined the impact of four photo-$z$ algorithms on the conclusions of the first DES cosmological analysis based on weak lensing discussed by \cite{Abb16a}. 

Photo-$z$ estimation will only get more challenging for the next DES releases and future photometric surveys. The reason is that we are sampling magnitudes beyond the reach of most spectroscopic surveys and therefore, traditional photo-$z$ validations are not realistic. This issue has inspired the implementation of new ideas in the collaboration, such as the calibration of photo-$z$s with cross--correlations \citep{New08,Davis17,Gatti18}, the training and validation of photo-$z$ codes with simulations (data--augmentation) \citep{Hoy15b} and validation of photo-$z$s with multi--band photometric samples \citep{Hoy17}. Techniques for assignment and validation of photo-$z$s for DES are under continuous development.

There are a large number of methods and algorithms available in the literature to compute and validate photo-$z$s. Thus, it is useful to work in an integrated environment where one can perform repeated tests and compare the results, while keeping the history well documented. Such an environment should provide the necessary hardware and software infrastructure to make feasible the comparison of different methods applied to large datasets. 

Besides dealing with big data, another remarkable aspect of current and near--future surveys is a large number of people working collaboratively. The computational methods are developed jointly by groups of people, commonly located in different countries. Therefore, it is useful to share a development environment that organizes software with version control, keeps the history, and ensures it is possible to reproduce results at any time.

Other web--based interfaces for astrophysical data mining and analysis are also being developed \citep[e.g., the DAMEWERE environment by][]{Bre14} aiming at the exploitation of large datasets.

The DES collaboration proposed, along with the Data Management system {\citep[DESDM\footnote{\url{http://www.darkenergysurvey.org/the-des-project/survey-and-operations/data-management/}},][]{Moh12}}, the creation of a dedicated portal to solve some of the problems associated with the data processing.  This concept became the DES Science Portal, hereafter ``the Portal''. 

During the early days of the DES project, the Portal was conceived as an ``end--to--end'' (E2E) process where the data flowed through a chain of tasks to prepare science--ready catalogs and perform scientific analyses. Since then the Portal has undergone several implementations for various scientific goals. The complexity of the system has been growing accordingly to accomplish the science demands. Now, there are instances of the Portal at Cerro Tololo Inter--American Observatory (CTIO), at the National Center for Supercomputing Applications (NCSA) and, at the Laborat\'orio Interinstitucional de e--Astronomia (LIneA)\footnote{\url {http://www.linea.gov.br/}}. In this paper, we refer to the instance at LIneA as ``the Portal''. 

The Portal provides the infrastructure necessary to handle large amounts of data, a common demand in extragalactic astronomy, but also attacks specific needs of the DES science, for instance, creating and applying systematic maps, computing zero-point corrections, performing star-galaxy classification, computing photo-$z$s and galaxy properties. The Portal generates galaxy samples in the form of pruned lightweight catalogs containing only the columns required by specific science analysis, which may also be integrated into workflows \citep{Fau18}. 

In this paper, we present, in particular, the capabilities of the Portal to produce photo-$z$s. It provides an integrated environment where all the steps necessary to compute photo-$z$s can be carried out in a controlled and consistent way. The automatic provenance, configuration management, and the computing facilities that sustain the Portal allow for a selection of many photo-$z$ algorithms or settings, which would be highly time-consuming without infrastructure such as this. The need for the Portal capabilities will increase as the DES databases grow, and more generally, as we enter an era of big data astronomy. 

In \citet{Cav15}, the authors of the PhotoRApToR algorithm discussed the advantage of linking automatically different steps of photo-$z$ calculation. The Portal surpasses PhotoRApToR in the sense that it is “method agnostic”: any photo-$z$ algorithm can be incorporated into the portal framework, which becomes especially interesting when the investigation aims to compare results using different methods.   

We present a sequence of tasks that include the preparation of a spectroscopic sample by combining data from different redshift surveys, the creation of training sets, the training and validation procedures for several algorithms, and the computation of photo-$z$s for large datasets. To show these examples, we used the DES first year data release, referred as Y1A1 \citep{Drl18, Abb18}.

The outline of this paper is as follows. In Section~\ref{sec:portal} we present the general technical aspects of the Portal. In Section~\ref{sec:pz_pipe} we go deeper in details of the processes related to the production of photo-$z$s. Still in Section~\ref{sec:pz_pipe} we present a use case of how the Portal can aid to determine reliable photo-$z$s through examples of runs using data from DES. The data is described in \ref{app_sec:data}. Finally, our conclusions and a summary of the paper are presented in Section~\ref{sec:conclusions}.

Also, we present, attached to this text, a list\footnote{\url{https://www.youtube.com/playlist?list=PLGFEWqwqBauBIYa8H6KnZ4d-5ytM59vG2}} of five videos (V0 to V4), showing examples of live runs, in a guided tour through the photo-$z$ production on the Portal.  

    \section{The DES Science Portal: Overview}
    \label{sec:portal}

Before describing the technical aspects of the Portal, we define in Table~\ref{tab:glossary} a list of terms frequently used in this text.

\begin{table}
    \small
    \begin{center}
    \caption{Glossary of terms used in the description of the Portal.}
    \begin{tabular}{|p{1.4cm}|p{6.4cm}|}
\hline
\ \ \ \ \ Term & \ \ \ \ \ \ \ \ \ \ \ \ \ \ \ \ \ \ \ \ \ \ \ \ \ \ \ \ \ \ \  Meaning \\
\hline
 Component & A Python script that works as a module to perform a specific task or to serve as a wrapper for an external algorithm. \\
\hline 
 Pipeline & A self--consistent sequence of components defined in an XML file. The pipeline script also defines dependencies between different components, as well as their required inputs and outputs. \\
\hline
 Workflow & One or a group of pipelines oriented to a common purpose. In general, it refers to scientific pipelines (known as Science Workflows).  \\
\hline
 Class~of products& A unique name that defines the attributes, characteristics and possible applications of a dataset or a product created in the Portal. \\ 
\hline
 End--to--end (E2E) & Sequence or chain of pipelines running in the Portal, starting with data acquisition, passing through several steps of data preparation and estimation of value--added quantities, culminating in the production of science--ready catalogs (see ). \\ 
\hline
 Stage & A group of pipelines in the same phase of data management. The E2E comprises the three stages: Data Installation $>$ Data Preparation $>$ Catalog Creation, and it is directly connected to the Science Workflows' stage. This last stage comprises the suite of scientific pipelines to support the research done by members of DES--Brazil Consortium (details in the supplemental video V0\footnote{\url{https://youtu.be/9zy0vXAWUdU}}). \\ 
\hline
    \end{tabular}
    \label{tab:glossary}
    
    \end{center}
\end{table}

The Portal, which the LIneA team designed, developed, hosts, operates and maintains, is an overarching web--based system solution for many issues faced by large astronomical surveys. Geographically, the operation is divided into three Portal instances running independently at CTIO, LIneA, and NCSA, as illustrated in Figure~\ref{fig:flow_portal_inst}. Each instance is responsible for accomplishing tasks in distinct phases of the data lifecycle. 

In the very beginning of the production of raw data, still at the telescope, the Portal{@}CTIO runs a pipeline called \textit{\textbf{Quick Reduce}}. It performs a rapid inspection of images, immediately after the exposures are taken, to detect possible problems and to produce a preliminary quality assessment. DESDM at NCSA receives the data and reduces and co--adds the raw images \citep{Mor16}. The photometry tables of objects detected from the coadded images are called \textit{coadd tables}. After that, these tables are downloaded and ingested into the Portal{@}LIneA's database, where the science--ready catalogs are created. Finally, the catalogs and science products are transferred to Portal{@}NCSA, which provides visualization tools through the data release interface.

In this work, we only present the technical aspects of the Portal{@}LIneA, which are related to the production of photo-$z$s. Among these features, we highlight those that apply to many other tasks: 
\begin{itemize}
\item Storing and registering of survey data to serve as input for analysis pipelines. 
\item Maintaining analysis codes (and their development history) ready to run on registered products.
\item Integrating external scientific codes publicly available into science workflows.
\item Facilitating the run and scalability of algorithms for scientific analysis (user--specific or collaboration--defined).
\item Registering outputs as products available for download and making them usable as inputs for other codes.
\item Keeping track of provenance (inputs, selected options, version of codes, etc.)
\item Allowing reproducibility of analysis results by keeping documentation about data and codes used, and operations performed.

\end{itemize}

\begin{figure}
    \includegraphics[width=1.\columnwidth]{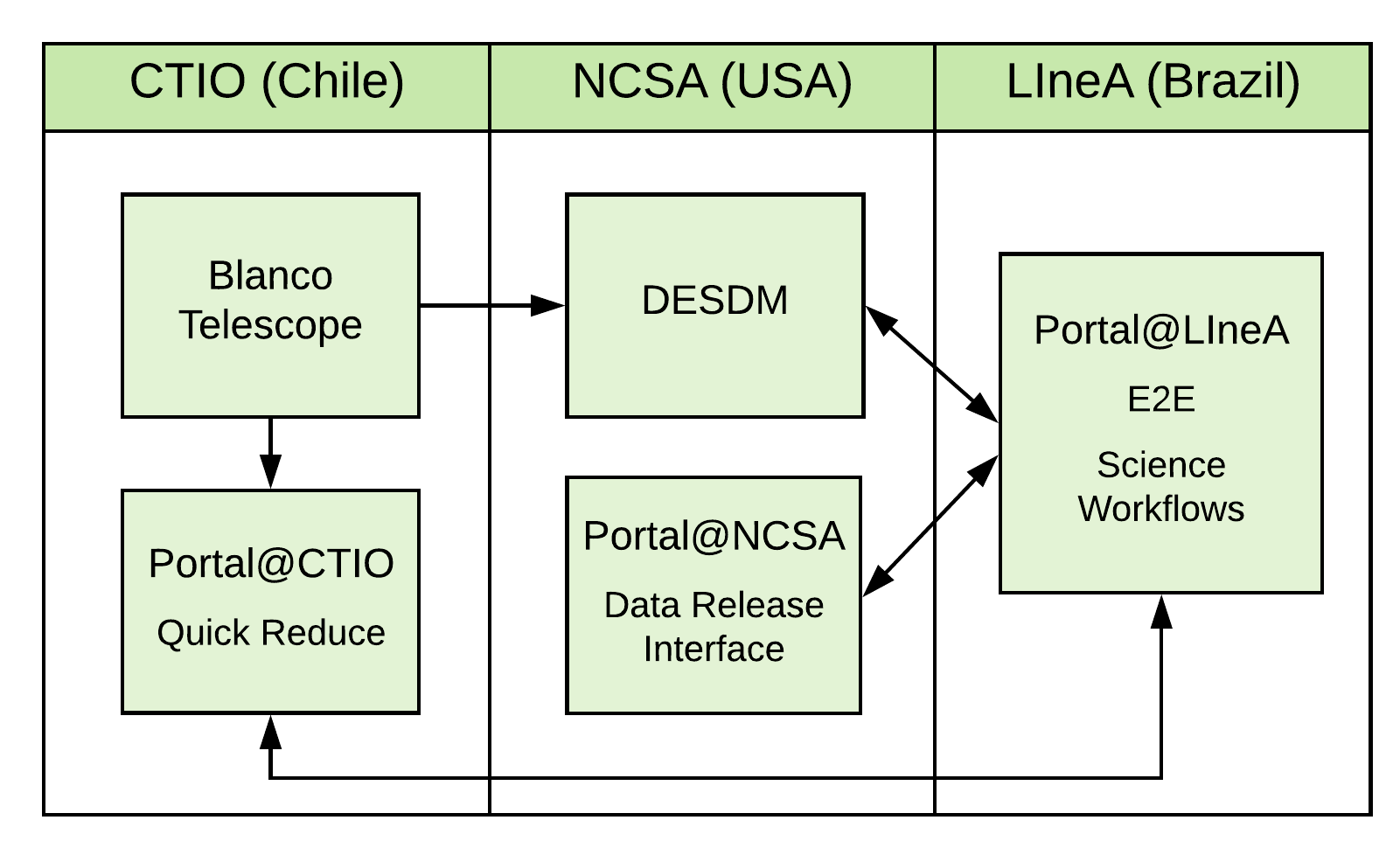}
    \caption{Instances of the DES Science Portal. The arrows indicate the data flow from the Blanco Telescope, at CTIO, through the various portal instances and the DESDM system at NCSA.}   
    \label{fig:flow_portal_inst}
\end{figure}

    \subsection*{Portal Infrastructure}


The Portal framework relies on two databases, a mass storage file system, a web interface, a workflow system and a cluster of computers, as illustrated in Figure~\ref{fig:portal_infra}.

Both databases uses PostgresSQL\footnote{\url https://www.postgresql.org/} object-relational database management system. The catalog database stores the catalogs retrieved from NCSA database and the catalogs produced by the Portal pipelines. The total storage capacity of the machine available for the catalog database is 23 TB, from which, $\sim$17 TB is already occupied. This device has a hot backup duplication in another connected machine. Both will be replaced in the future by new devices with larger capacity.

The administrative database keeps track of metadata such as available releases, ingested products, product information, such as file and table names, storage location, classification, provenance, etc. All the operations and steps are logged in the administrative database, so it allows to detect errors and investigate them posteriorly, and also to produce reports on the resources usage that can be filtered by a user or by an application.

The mass storage device has the capacity for 59 TB, from which 39 TB is already used. It keeps data in three separate spaces: 

\begin{itemize}
\item \textbf{Archive} area, where the original catalogs FITS files are preserved (so the catalogs are duplicated in the database and mass storage's archive). 
\item \textbf{Scratch} area, where are placed the directories and files that are created during the process executions. They include tables, images, flat files of any kind, run logs, error logs, etc. They remain in this area until the user saves the process. Periodically, a tool called \textit{garbage collector} removes old files from this area. 
\item \textbf{Process} area, is a safe area where the directories and files from saved processes are kept permanently.   
\end{itemize}

The location of all the process directories, both in Scratch and Process areas are registered in the administrative database.

\color{black}


The web application front--end uses the Model-View-Con\-trol\-ler \citep[MVC,][]{Bur87} software architecture and is developed in both Hypertext Markup Language\footnote{\url{ https://www.w3.org/html/} } (HTML) and Java\-Script \footnote{\url {https://www.javascript.com/} } languages. It connects to the databases via the back-end infrastructure Python components. 

\begin{figure*}
    \begin{center}
    \includegraphics[width=1.7\columnwidth]{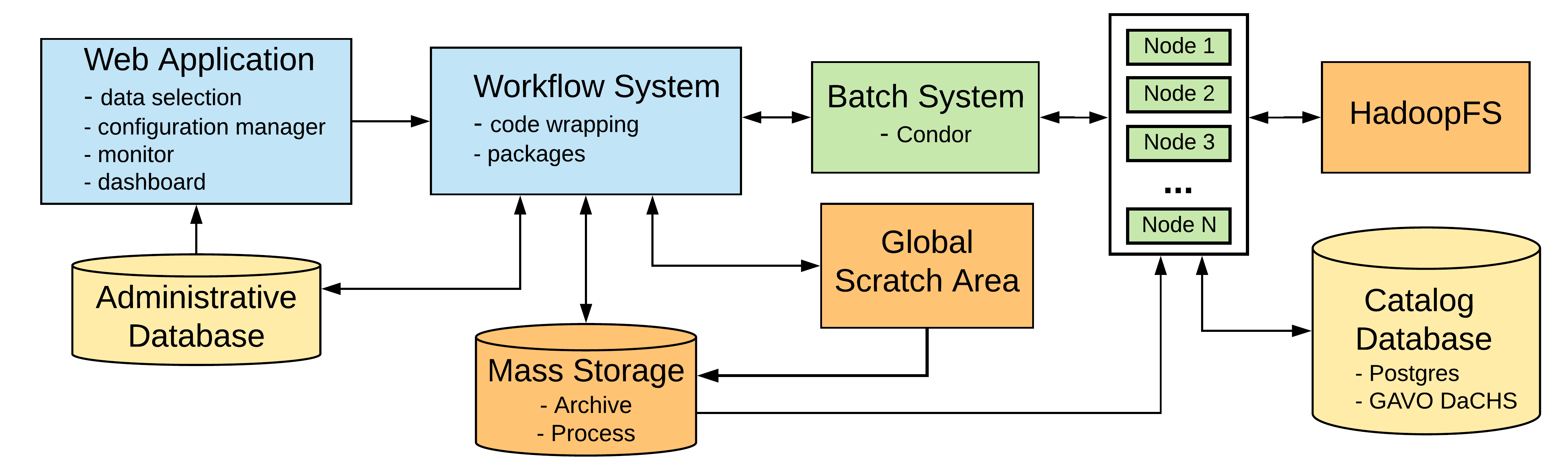}
    \caption{Elements of Portal's infrastructure: software components (in blue), processing systems (in green), and storage systems (databases in yellow, file systems in orange). }
    \label{fig:portal_infra}
    \end{center}
\end{figure*}

   \subsection*{Pipelines and Components}

The computational tools available in the Portal are organized in pipelines and components. The former are workflows defined in Extensible Markup Language \citep[XML,][]{bray2008extensible} which concatenates a chain of tasks performed by components. They determine the order of execution of the tasks and the parallelization strategy, as well as necessary inputs and outputs. 

Components are Python scripts which can both serve as a wrapper for an external code or be an independent algorithm. Scientific codes, which can be written in any programming language, are encapsulated by the wrappers,  which are in charge of preparing the inputs in the format expected by the code, calling the code to run, and handling the outputs.

\begin{figure}
    \begin{center}
    \includegraphics[width=\columnwidth]{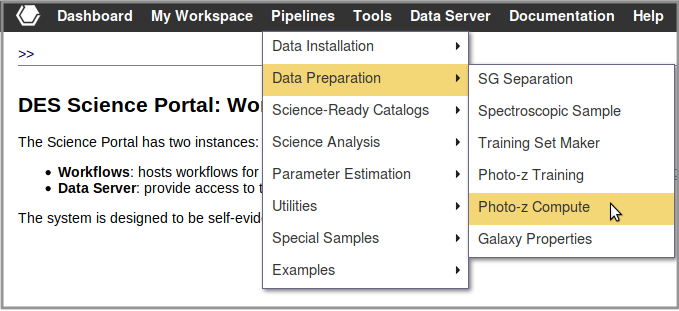}
    \caption{Portal's initial screen - data preparation pipelines menu.}
    \label{fig:pipeline_menu}
    \end{center}
\end{figure}

The pipelines are triggered via the Portal interface (see Figure~\ref{fig:pipeline_menu}), where the user navigates through tabs to define inputs and configuration parameters.  {For each pipeline, there is a README document that includes configuration tips and pieces of advice regarding the technical aspects of pipeline running. There are also cookbooks with a scientific approach to help the user on decisions about data selection and data-dependent configurations.}

When a process finishes, the user receives a notification via e--mail with a link to a product log -- a page containing results and process--relevant pieces of information.

The pipelines are self--consistent independent building blocks in the E2E chain. Each one provides a product log and can be redone as many times as the user demands. {The list of the most used pipelines can be seen in Figure~\ref{fig:E2E_complete}.} The pipelines highlighted in gray are those related to the photo-$z$ calculation. They will be discussed in details in Section~\ref{sec:pz_pipe}. For now, it is not possible to run part of a pipeline stand--alone. For instance, one can perform repeated tests, varying configurations, until it converges into a result with the desired quality. In our case of study, we have already run training, validation, and computing for nine different photo-$z$ codes, varying their configurations, more than 300 times in total, considering the several datasets of the Y1A1 release. That would be complicated to manage using directories and command--line runs.

\begin{landscape}
\begin{figure}
\begin{center}
    \includegraphics[width=0.9\columnwidth]{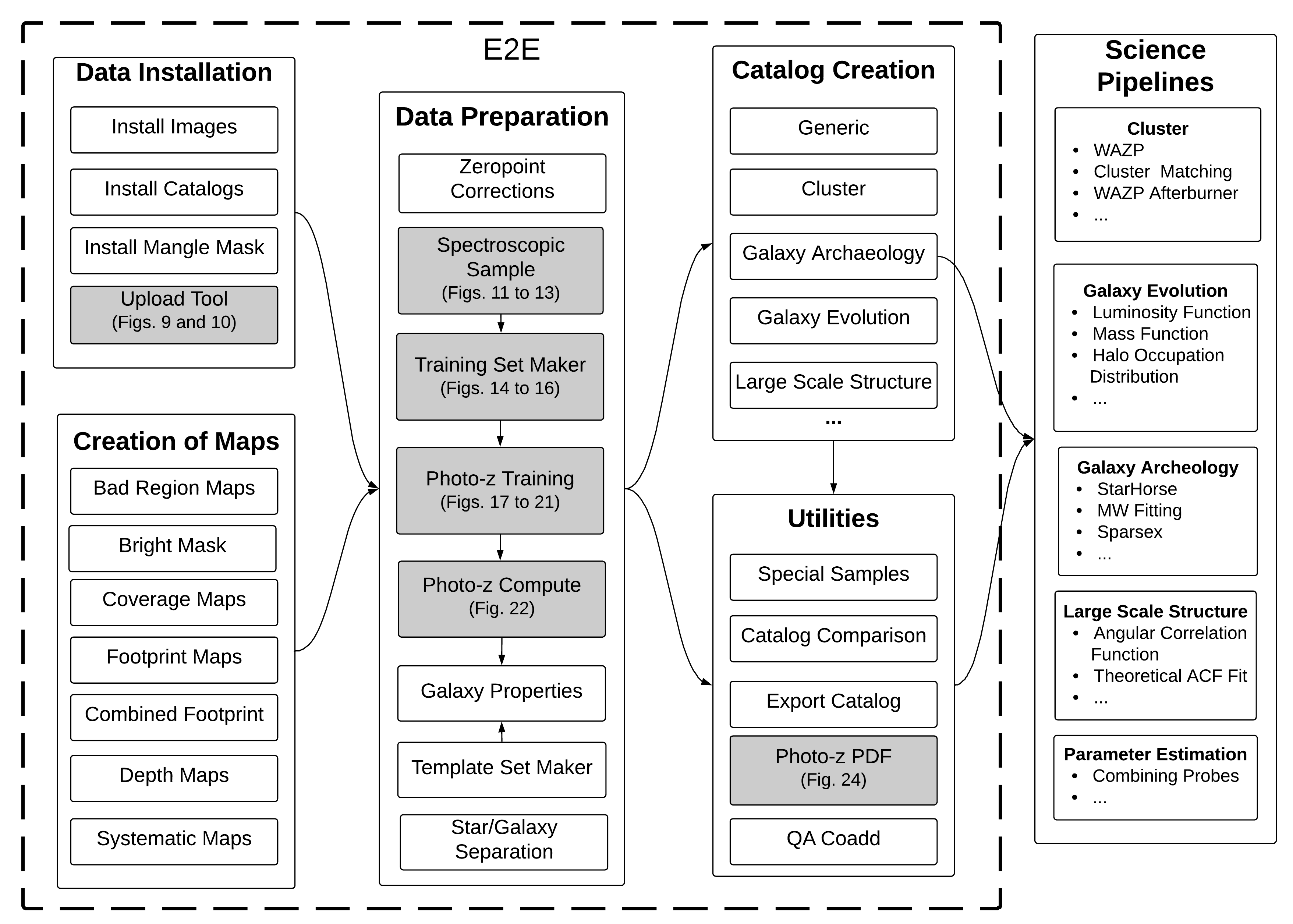}
    \caption{Sequence of pipelines organized by stages. The data flow mainly from left to right. The pipelines inside the dashed line belong to the E2E, where science-ready catalogs are produced and delivered to the Science Workflows. The pipelines highlighted in gray are those related to the photo-$z$ calculation. The figures indicated inside the parenthesis are the respective screenshots.}
    \label{fig:E2E_complete}
\end{center}
\end{figure}
\end{landscape}


The Portal is developed collaboratively using the GIT\footnote{\url{https://git-scm.com/}} version control system. All changes done by different developers are merged and intensively tested in a separate Portal clone called ``Testing'', to ensure consistency and compatibility between developers' versions (see Figure~\ref{fig:portal_clones}). The stable and validated versions of the codes are deployed to the production Portal. All the technical information about hardware mentioned in this work refers to the production environment. 

Pipelines and component codes are open\-source\footnote{\url{https://git.linea.gov.br}}. These codes often have dependencies on other programs or libraries which are provided by two systems: Tawala and the Extended Unix Product System (EUPS\footnote{\url{https://github.com/RobertLuptonTheGood/eups}}). Tawala is a homemade repository system created in earlier stages of portal development, which we maintain due to the large number of historical code dependencies. Nowadays it is kept frozen and is being gradually replaced by EUPS. 

\begin{figure}
    \begin{center}
    \includegraphics[width=\columnwidth]{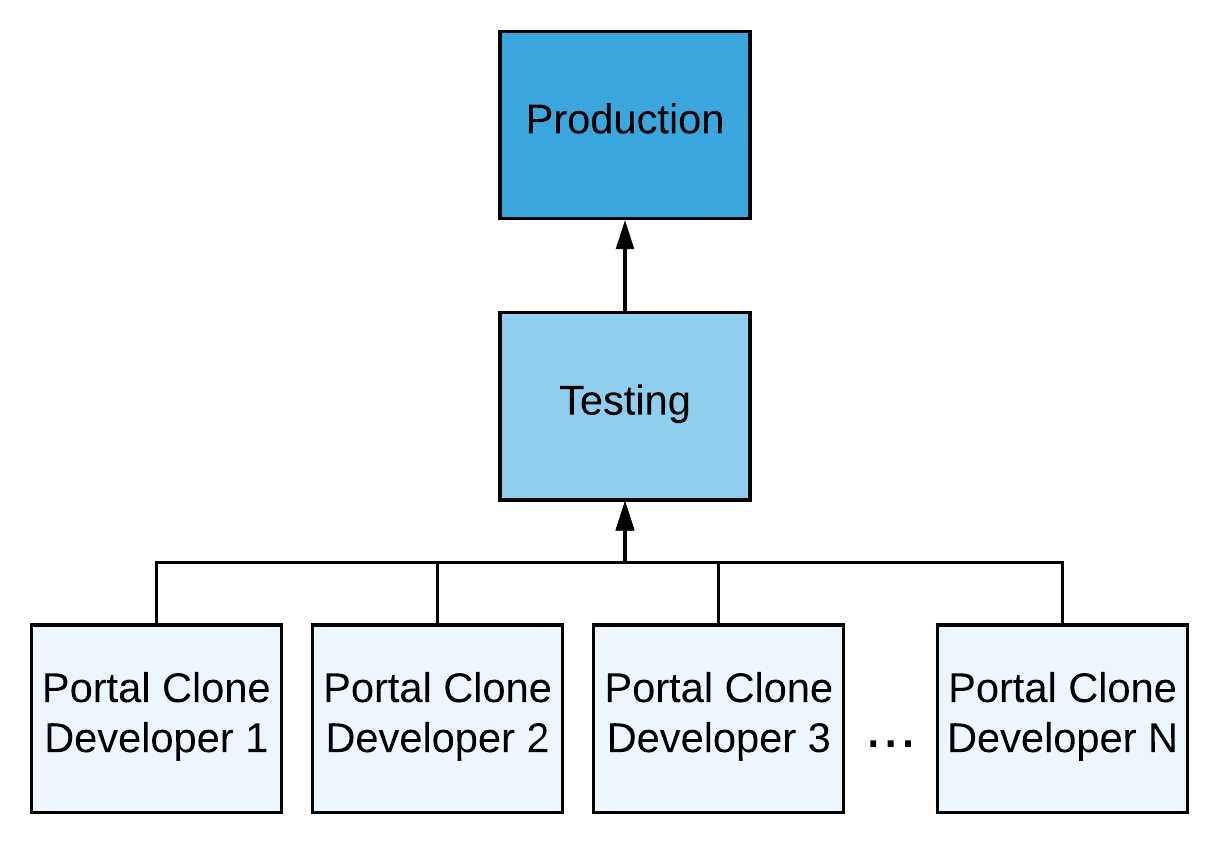}
    \caption{Portal environments: development (bottom), testing (middle), and production (top).}  
    \label{fig:portal_clones}
    \end{center}
\end{figure}

{The pipelines of the first stage, \textit{\textbf{Data Installation}} (the first block of pipelines in Figure~\ref{fig:E2E_complete}), are only executed by the portal developers (LIneA's IT team), except for the \textit{\textbf{Upload Tool}}, that can be executed by any science user. The initial step of \textit{\textbf{Data Installation}} is the} retrieval and ingestion from NCSA, of \textit{coadd tables}. A daemon process called \textit{\textbf{Data Retriever}} periodically inspects the NCSA database to discover new releases of the DES Survey data. For each new release recognized, the \textit{\textbf{Data Retriever}} registers its existence in the Portal Administrative database. At this point, the Portal system ``knows" the presence of a new release and a Portal operator can start the specific installation procedure that takes care of downloading the data in an optimized network route. The download is done using a server in a ``demilitarized zone network'' and then storing it into a mass storage server. Later it is ingested into the Portal Catalog database. Finally, the Administrative database registers the new tables and makes them available to be accessed by visualization tools and to serve as input to be processed by pipelines in the Portal.

A system of ``classes'' of products connects pipelines via inputs and outputs. In this context, a class is a unique keyword that identifies a specific product data structure. Hence, we can define the type of products that each pipeline receives as input and returns as output. {As an example, the catalogs mentioned above, which are ready to be used in the portal, pass through the pipeline \textit{\textbf{Install Catalog}}, where they are classified as products with class =``Object Catalog''. This way, they are made available to each pipeline where we set this class as input.} \textit{\textbf{Star/Galaxy Separation}} is one of the pipelines configured to use products of this class as input, so when the user chooses to run this pipeline, the system will display the tables registered by \textit{\textbf{Install Catalog}}, and the user will be asked to select one of them. Then, running various pipelines can be understood as a hierarchy of products, as shown in Figure~\ref{fig:provenance}. This figure illustrates the provenance of a sequence of pipelines related to photo-$z$ pipelines described in Section~\ref{sec:pz_pipe}.

    \subsection*{Parallelization}
    \label{subsec:parallel}

The photo-$z$s estimation is one of the most computationally intensive tasks of the E2E process, due to the size of datasets involved. The first solution adopted in the Portal is the so-called \textit{Embarrassing Parallelization} \citep{her11},  which is the division of the data into small partitions and their processing is done simultaneously by several computers.  

The large volume of data requires a high--performance system to transfer such data inside and outside each computer node avoiding the creation of an I/O bottleneck. This is a common problem in data--intensive computing that does not have a unique solution for all possible use cases. We solved this obstacle by implementing different I/O service strategies that can be used according to each specific problem, as detailed below. 

\begin{figure}
    \begin{center}
    \includegraphics[width=1.0\columnwidth]{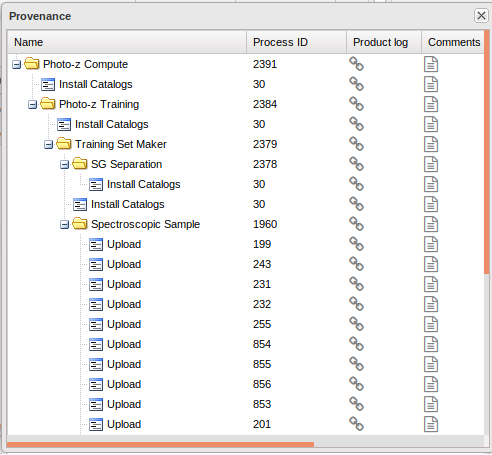}
    \caption{Display of provenance chain {\bf for} a product of Photo-$z$ Compute pipeline (first column). Each sub--level tag the processes that entered the parent process with their identification number (process ID) and their links to the product logs and comments made by users. }   
    \label{fig:provenance}
    \end{center}
\end{figure}

For the very basic DES datasets (table of objects identified in coadded images) frequently used in the majority of the algorithms, we use the high responsive Hadoop Distributed File System \citep[HDFS,][]{shvachko2010hadoop} to distribute data uniformly across the cluster nodes. Despite the fact that the processing is done by several computers, the data itself also needs to be distributed. Otherwise, the simultaneous reading from the same database by several parallel tasks (jobs) would establish a significant bottleneck.  

The Portal's computers cluster contains 38 nodes with 24 central processing unit (CPU) cores each. The management of job submission is performed by an orchestration system, together with the HTCondor management system\footnote{\url{https://research.cs.wisc.edu/htcondor}}. The orchestration system interprets the parallelization strategy defined by each pipeline and its configuration, then it calculates the number of jobs necessary, and gives instructions to HTCondor. The tasks are organized in the cluster nodes based on the data, such that it always prioritizes the runs to process the data that are already stored in each node, avoiding unnecessary data transfer overheads. If necessary, it allows reading additional data from other nodes. There is a mirroring in the data storage to help on this optimization. Each data chunk is stored triplicated in three different nodes. Hence, if one node needs data from a neighbor node that is, by chance, very busy with some intense process, the former still has two other options of nodes from where to get the data. 

For less frequently accessed data used in our algorithms, the Portal use PostgreSQL\footnote{\url{https://www.postgresql.org/docs/9.6/static/parallel-query.html}}, a fat node database with 24 cores and 256 gigabytes of RAM that supports multiple queries concurrently, allowing the fast retrieval of the portion of the data that each node needs. Moreover, to fast retrieve positional data according to its spatial position, we use the Q3C \citep{Kop06} PostgreSQL extension for spatial indexing on a sphere. 

For temporary data, such as the one produced in a component that needs to be consumed by the next segment of the same pipeline, it is required to be staged and then rapidly transferred from one or more nodes to others. In this case, we use the high--performance parallel Lustre file system \citep{donovan2003lustre}, explicitly developed for large--scale cluster computing. Finally, PostgreSQL is used to store the new generated products temporarily stored in a Network File System \citep[NFS,][]{sandberg1985design}. An example of such a product is the product log generated by any pipeline.

\begin{figure}
    \begin{center}
    \includegraphics[width=1.\columnwidth]{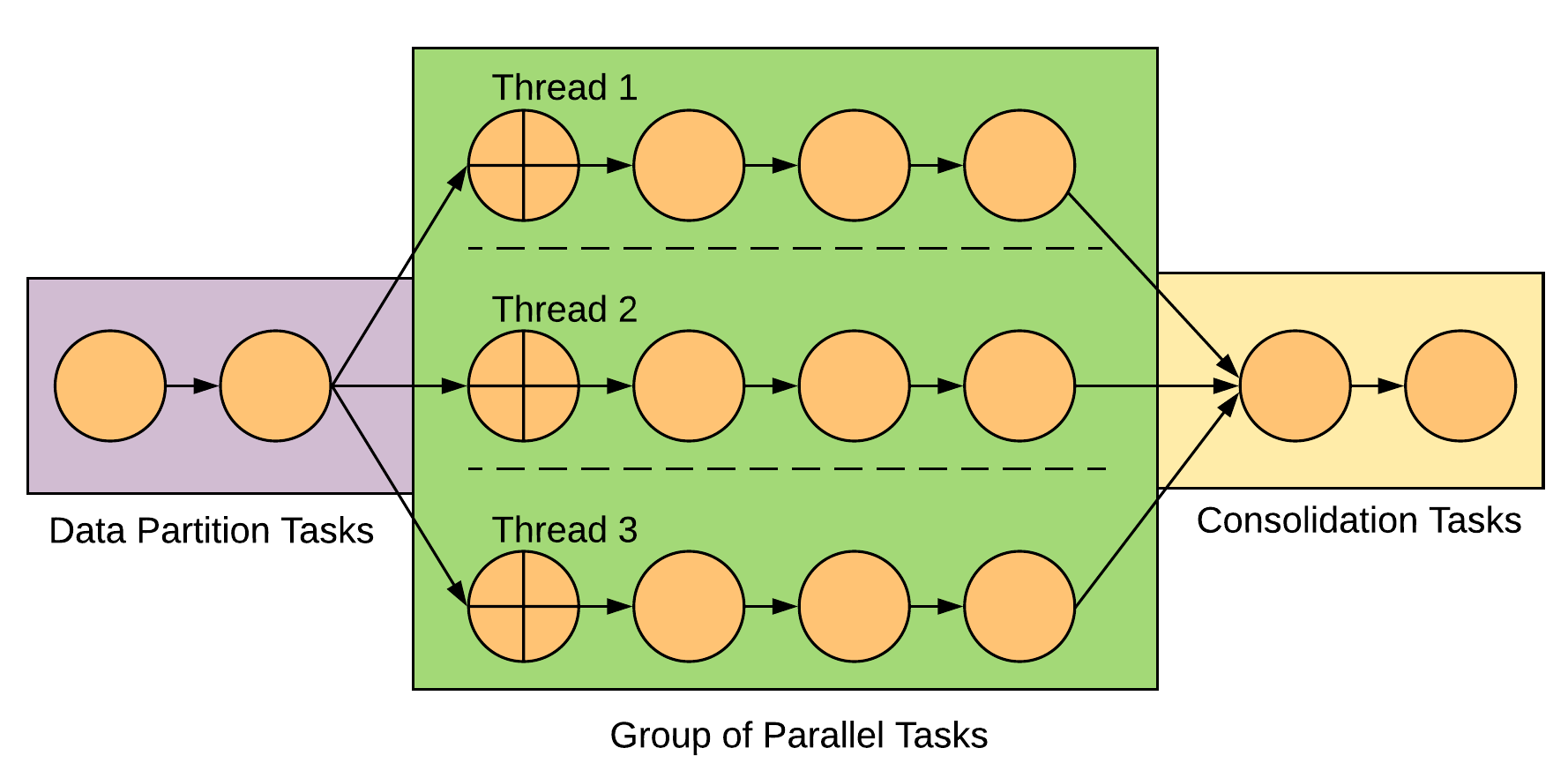}
    \caption{Illustration of parallel processing managed by the workflow system. The first group of tasks defines the data partitioning schema and distributes jobs to the cluster. Tasks that require the same chunk of data are grouped in the same cluster node, to minimize data transfer. The first component of the task group running in parallel is responsible for data retrieving (marked with a cross symbol). The final task group consolidate the results from all jobs.}
    \label{fig:parallel}
    \end{center}
\end{figure}

The parallel processing is implemented in a ``Map-Reduce'' \citep{Dea04} way, as illustrated in Figure~\ref{fig:parallel}, and it follows one of the three options of data partitioning strategies:

\begin{itemize}
\item \textbf{Tiles}: This is the original data division from DES. Tiles are tangent projections of an array equivalent to 5000 $\times$ 5000 CCD pixels \citep[0.7306$^\circ$ on each side,][]{Mor16}. Each tile corresponds to one file stored in HDFS. The file size is strongly dependent on the density of objects detected, which is related to the depth of observations. For the Y1A1 depth, one tile contains, typically, $\sim 40$k objects detected, and the file occupies $\sim$130 MB (for data description, see~\ref{app_sec:data}). The number of tiles processed by each job submitted to the cluster is a free parameter in the configuration of the pipelines. The optimal choice depends on the dataset size (number of tiles), compared to the number of CPU cores available, as well as the memory consumption of the algorithm run. 

\item \textbf{HEALPix\footnote{\url{http://healpix.sourceforge.net/}} indexation}: This is a more flexible way of splitting the entire sky projected area into pieces (pixels) of variable size. The division based on HEALPix pixels is available both for the \textit{coadd tables}, stored in HDFS, and some of the other datasets stored in the database. Concerning the first, the data is distributed in files containing pixels of \textit{NSide} = 32 ($\sim$1.6 GB each for Y1A1 wide datasets), using nested pixel ordering. \textit{NSide} is a quantity that represents the resolution of the HEALPix map, so that the total number of pixels covering the whole sphere is $N=12 \times(\textit{NSide})^{2}$. Therefore, the larger the \textit{NSide}, the smaller the pixel size. Similarly to the tile--based partitioning, the choice of \textit{NSide} has implications on the processing performance. The available options for \textit{NSide} values are $2^{n}$, $n=$ 2 to 10. Stress tests have shown that large pixels ($n<4$) should be avoided due to random--access memory limitations, depending on the dataset. On the other hand, scaling--out to use small pixels ($n>8$) are also not recommended, because they convert in too many jobs, increasing the transferring overhead and stressing the cluster management system. One advantage of the partitioning based on HEALPix is that it applies to any data set which has celestial coordinates, therefore connected to spherical geometry. It is particularly useful to organize simulated data, that is not related to any observational strategy. 

\item \textbf{Custom}: This option defines the data partitioning based on one of the data attributes (i.e., any table's column). It queries the data and distributes, among the several cluster nodes, using intervals of one attribute. There are three options of binning: (i) fixed: evenly spaced bin edges; (ii) variable: evenly populated, so the bin widths are irregular; (iii) manual: bin intervals are defined manually by the user. This data partition is more commonly used by the science workflows. Some examples of this parallelization are the estimation of the angular correlation function of galaxies in tomographic bins of redshift, and the estimation of the luminosity function of galaxies in bins of absolute magnitude. 

\end{itemize}

In all cases, the scaling of parallel processing is not automatic. The user is asked to make decisions on the configuration screen about the size or the number of partitions. Optionally, the second layer of parallelization can be applied as a further user--specific parallelization, by reserving one or more machines and distributing the process among their cores using, e.g., Python Multiprocessing. All this flexibility is put in place to optimize the pipeline execution in various scenarios, depending on the peculiarities of each run. Too large data chunks can cause memory over--heading problems or take too long to be processed, while too small pieces can waste time with data transfer and create a massive queue of jobs waiting for available nodes. Therefore, the optimal parallelization schema must be defined case by case. In Section~\ref{subsec:pz_compute} we show, as an example, a test to measure the impact of the configuration chosen on the processing speed, measured by the processes total duration and the time spent in groups of components.

    \section{Photo-z Pipelines}
    \label{sec:pz_pipe}

The estimation of redshifts is a fundamental part of the process of creating science--ready catalogs for extragalactic applications. In photometric surveys, photo-$z$ methods and algorithms are used to surpass the lack of spectroscopic information. In most cases, the photo-$z$ estimation and validation rely on the use of a ``true'' sample, in the sense of assuming negligible errors in the determination of their redshifts, as in the spectroscopic redshift samples. This sample will be useful both to train empirical algorithms (e.g., neural networks, nearest--neighbor), and to estimate the uncertainties in the mean values and errors of their distributions. We note that as photometric surveys are reaching fainter magnitude limits, the spectroscopic data available are less representative of the photometric sample. Therefore, new techniques are being developed to estimate photo-$z$ in surveys without the need of spectroscopic data \citep[see, e.g.,][]{Hoy17}. In the  Portal, we have implemented tools to estimate the redshift of sources using standard techniques based on spectroscopic redshifts to train and validate the algorithms. 

In the following sections, we describe the methodology of each pipeline related to the photo-$z$ estimation in the Portal. They operate in sequence, as illustrated in Figure~\ref{fig:photoz_flowchart}. For each pipeline described, we add an example of a result obtained using data from DES Y1 release, as an illustration, and proof of concept. A description of the data used is available in \ref{app_sec:data}.

\begin{figure}
    \begin{center}
    \includegraphics[width=0.8\columnwidth]{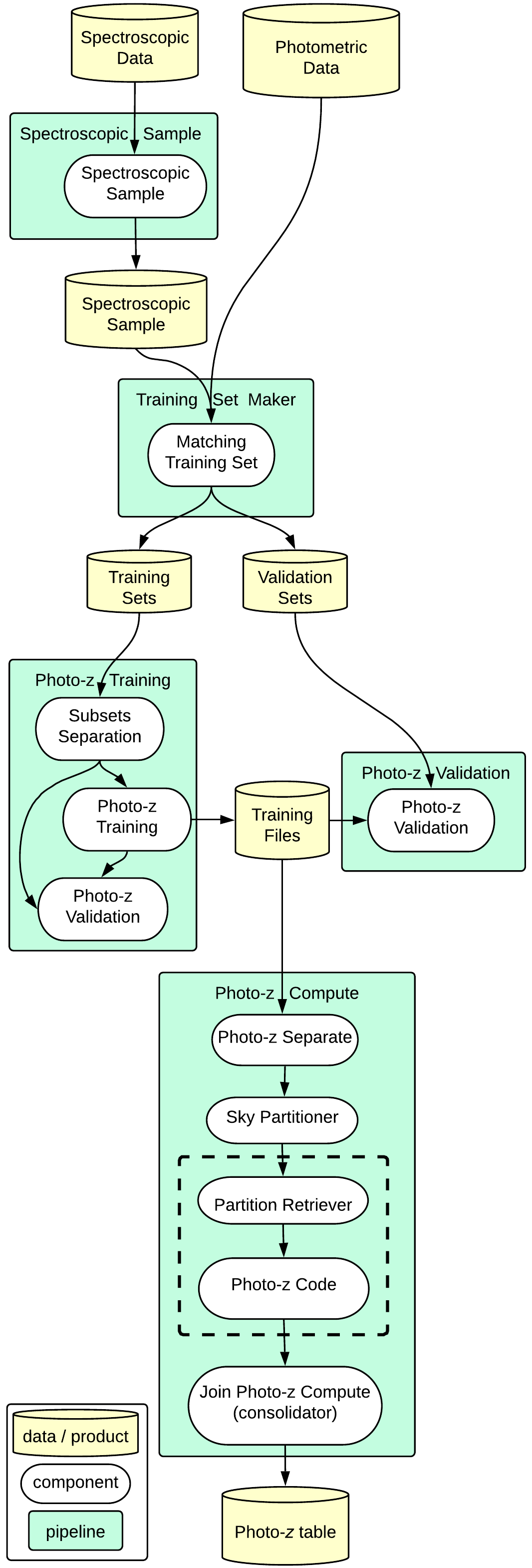}
    \caption{Photo-$z$ complete workflow. Pipelines are represented in green: \textit{\textbf{Spectroscopic Sample}} (Section~\ref{subsec:spec_sample}), \textit{\textbf{Training Set Maker}} (Section~\ref{subsec:ts_maker}), \textit{\textbf{Photo-z Training}} and \textit{\textbf{Photo-z Validation}} (Section~\ref{subsec:pz_train}), and \textit{\textbf{Photo-z Compute}} (Section~\ref{subsec:pz_compute}. There is also another pipeline (\textit{\textbf{Photo-z PDF}}, Section~\ref{subsec:pz_pdf}) not present in this illustration, because it is placed after the construction of scientific catalogs. The components are represented in white. Data inputs and products are represented by yellow cylinders. In \textit{\textbf{Photo-z Compute}}, the dashed line involves the part that runs in parallel. The generic ``Photo-z Code'' component block represents one of the components that wraps a particular photo-$z$ code (DNF, LePhare, etc). } 
    \label{fig:photoz_flowchart}
    \end{center}
\end{figure}

    \subsection{Spectroscopic Sample}
    \label{subsec:spec_sample}

In the Portal the first step to obtain photo-$z$s is the creation of a spectroscopic sample that will be matched with DES photometric catalog to define a training set used for training and validation. The goal is to create a sample with as many sources as possible avoiding effects of cosmic variance or under-re\-presenta\-tion of particular spectral types. 

The database associated with the Portal serves as a centralized spectroscopic database for DES \citep{Hoy17}, being continually updated, in particular, by ongoing follow-up observations from DES collaborators such as the OzDES project \citep{Yua15,Chi17}, as well as with substantial new spectroscopic galaxy samples made public. 

The current spectroscopic redshift samples available are indicated in Table~\ref{table:all_specs}. Once a new spectroscopic survey of interest is identified, the data is downloaded to our archive and ingested in the database to be accessible by the \textit{\textbf{Spectroscopic Sample}} pipeline. At this stage of the process, it is necessary to provide some predefined information to be associated with a given spectroscopic sample in the registration database (Figure~\ref{fig:upload1}).

\begin{figure*}
    \begin{center}
    \includegraphics[width=1.6\columnwidth]{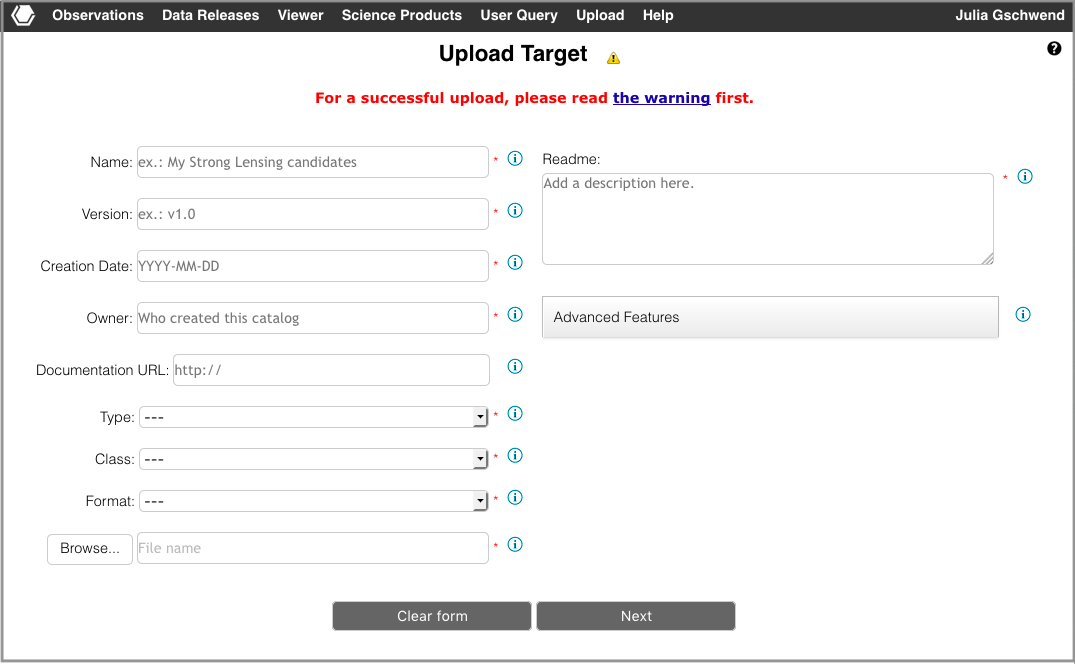}
    \caption{Upload tool initial screen. The user provides the relevant metadata to registering and versioning, such as the data source and a short description.}  
    \label{fig:upload1}
    \end{center}
\end{figure*}

\begin{figure*}
    \begin{center}
    \includegraphics[width=1.6\columnwidth]{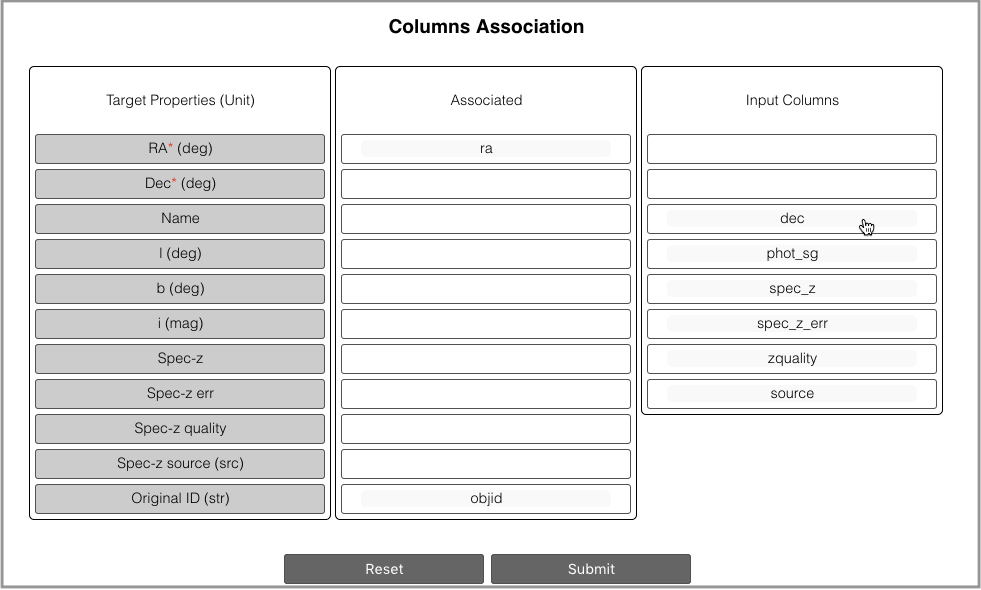}
    \caption{Upload tool - column association screen. In this screen, the user is asked to drag the original column names (on the right side) to those names expected by the pipelines (on the left). With this information, the \textit{\textbf{Upload}} tool creates a table in the database with the contents of the uploaded catalog and the columns names appropriate to be read by the pipeline \textit{\textbf{Spectroscopic Sample}}. Both the original and the ``translated'' names are saved as metadata, to keep the history.} 
    \label{fig:upload2}
    \end{center}
\end{figure*}

When building a spectroscopic sample from heterogeneous sources, we need to take into account the following: 
\begin{itemize}
\item Each source might have different column names for the same quantities, e.g., redshift represented by ``z'', ``spec-z'', ``zspec'', etc. We associate mandatory columns (RA, Dec, redshift, redshift quality, source, redshift error) when uploading each sample to ensure that the columns are properly delivered as input to the pipelines (see Figure~\ref{fig:upload2}). 

\item Each catalog may have different quality estimates of redshift. To normalize quality flags to a single schema, we take the approach of OzDES Survey \citep{Yua15}, with qualities ($Q_{spec}$) ranging from 0 to 4. The numbers 0 and 1 are two types of unknown redshift, 2 is only a guess, 3 is above 90\% confidence, and 4 attributed to a trusted redshift. When a survey is uploaded, we need to tell the centralized spectroscopic database that a new catalog has arrived and how to translate the quality information to the numerical system explained above. 

\item Elimination of duplicates. Internally, the pipeline handles possible multiple measurements for same objects. We implemented the following set of possibilities: 1) Select the spectroscopic redshift according to the best quality (default); 2) Make an average of all values to give a single redshift. To identify measurements for the same objects, the matching is done based on the angular separation between the coordinates in the different surveys using a matching radius of 1.0 arcsec (default). If two or more objects are within the search radius, the criteria to solve duplicates selected in the configuration tab applied is the following: we select the measurement that has the highest $Q_{spec}$. If more than one observation has the same $Q_{spec}$ flag, we select the one obtained more recently. If there are two or more observations in the same year, we choose the redshift with the smallest error, when it is available. Finally, if we still have more than one source (from the same year and with no errors available), we choose the one with redshift the closest as possible to the mean value of all the multiple measurements. Since we have selected a high ``quality'' threshold, the differences between choosing the best source or averaging between all the different matches are negligible. This step is done in several sub--steps using a \textsc{PostgreSQL} extension for spatial indexing on a sphere, called Q3C \citep{Kop06} and the Starlink Tables Infrastructure Library Tool Set \citep[STILTS,][]{Tay06}.  

\item Spectral type classification. Some surveys classify the source (star, galaxy, QSO). When this is available, we use this information to allow for specific spectral classification of stars or galaxies. In the case where no classification provided, we assign as ``stars" every object with $z<$0.001. For the time being, we do not classify objects as QSOs, and they will normally enter as galaxies. However, in the pipeline, \textit{\textbf{Training Set Maker}}, where we have photometric data associated to each object, we can apply a point source removal criterion, which eliminates most of the QSOs from the galaxy samples we built. 
\end{itemize}

\begin{figure*}
    \begin{center}
    \includegraphics[width=1.6\columnwidth]{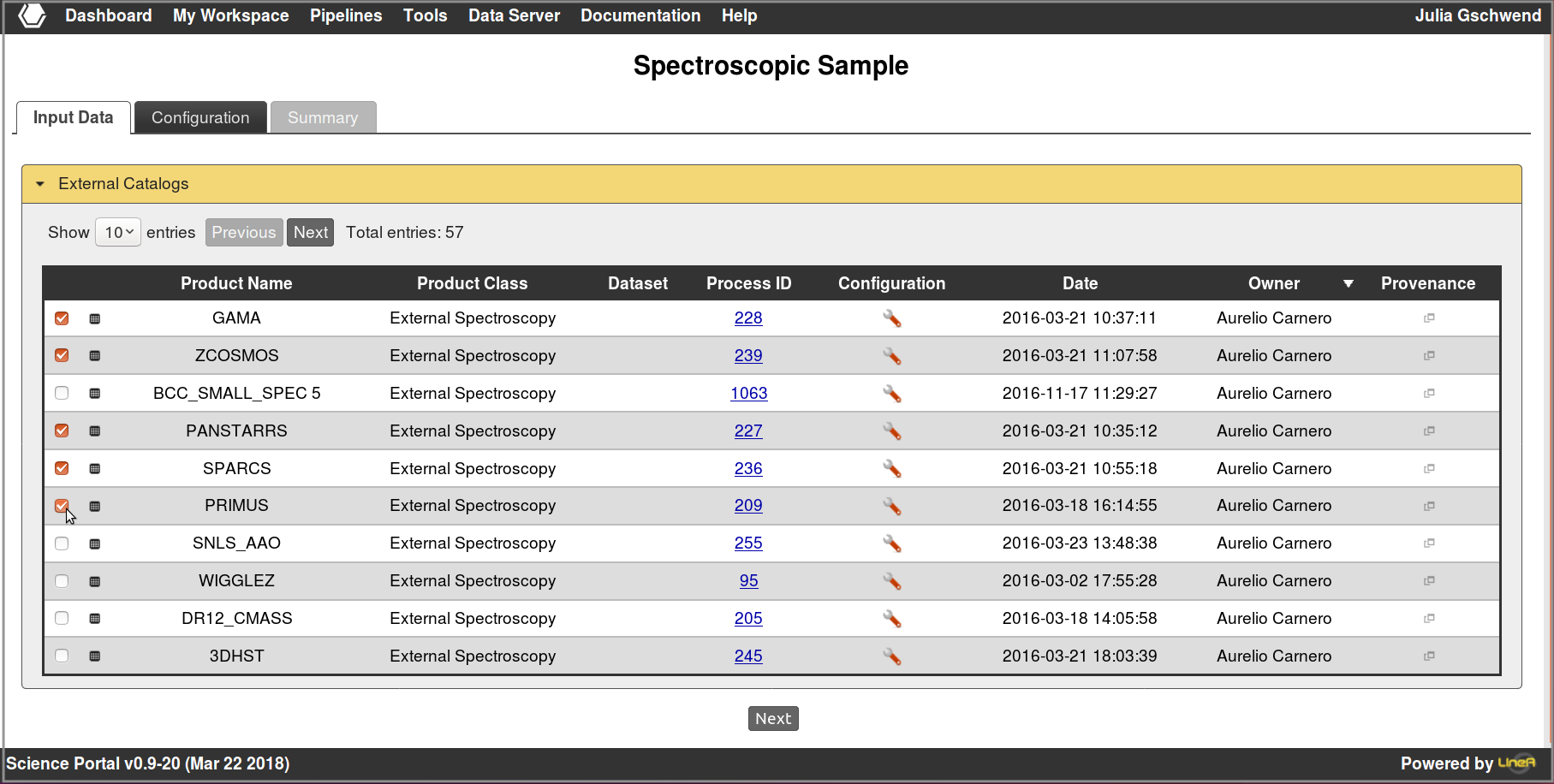}
    \caption{Input menu of \textit{\textbf{Spectroscopic Sample}} pipeline. In this tab, the user selects which spectroscopic surveys are to be include in the spectroscopic sample. }
    \label{fig:spec_input}
    \end{center}
\end{figure*}

    \begin{figure*}
    \begin{center}
    \includegraphics[width=1.6\columnwidth]{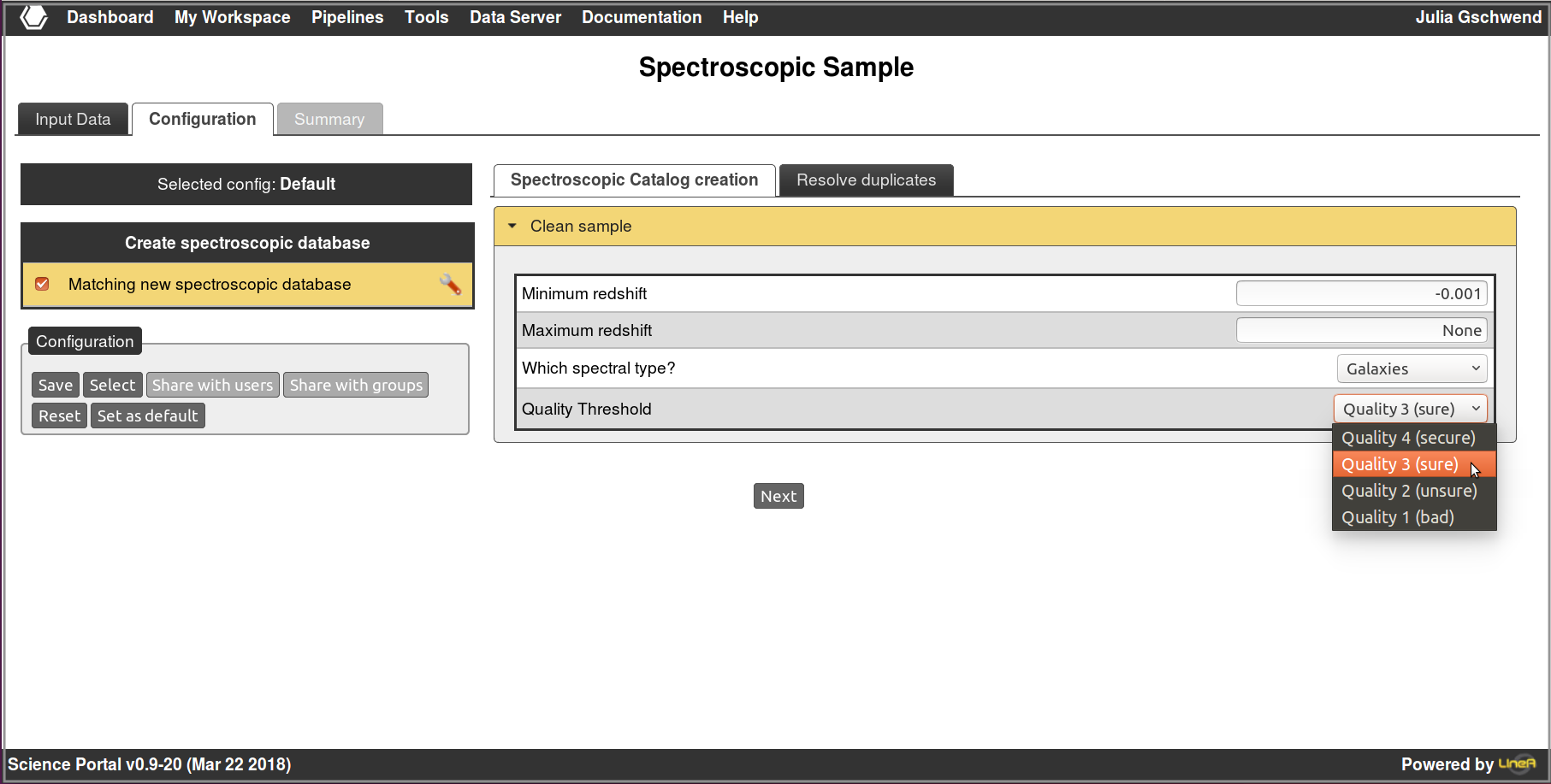}
    \end{center}
    \caption{Configuration menu of \textit{\textbf{Spectroscopic Sample}} pipeline. In this tab, the user selects spec-$z$ and quality criterion to resolve duplicates.}
    \label{fig:spec_config}
    \end{figure*}

In summary, the \textit{\textbf{Spectroscopic Sample}} pipeline  creates a ``master'' sample by combining spectroscopic data from various surveys. The user chooses the ones to include in a check--box menu (see Figure~\ref{fig:spec_input}) on the pipeline interface. The next step is to choose the configuration parameters, as can be seen in Figure~\ref{fig:spec_config}. There one can make decisions about the criterion adopted to handle duplicates, quality threshold, spectral classification, etc. 

The result of this pipeline is a table registered in the database containing: spec-$z$s, errors, quality flags, sources, and coordinates. At this stage, there is no association with DES objects. This table becomes available to be an input for the next pipeline or to be delivered to the collaboration. 

Relevant information about the spectroscopic sample created in a particular run is in the process' product log. In Figure~\ref{fig:spec_prodlog} we show the result of creating a spectroscopic sample containing data from all surveys available in the Portal, selecting only the best measurements ($Q_{spec}$=4), and resolving duplicates by the default criteria mentioned above. This plot shows the spatial distribution of spectroscopic redshifts included in the sample, and we can verify that part of the spectroscopic sample spills over the DES footprint. The product created contains redshifts of 1,408,138 unique galaxies, from 34 surveys (see \ref{app_sub:spec_data}).

    \begin{figure*}
    \begin{center}
    \includegraphics[width=1.3\columnwidth]{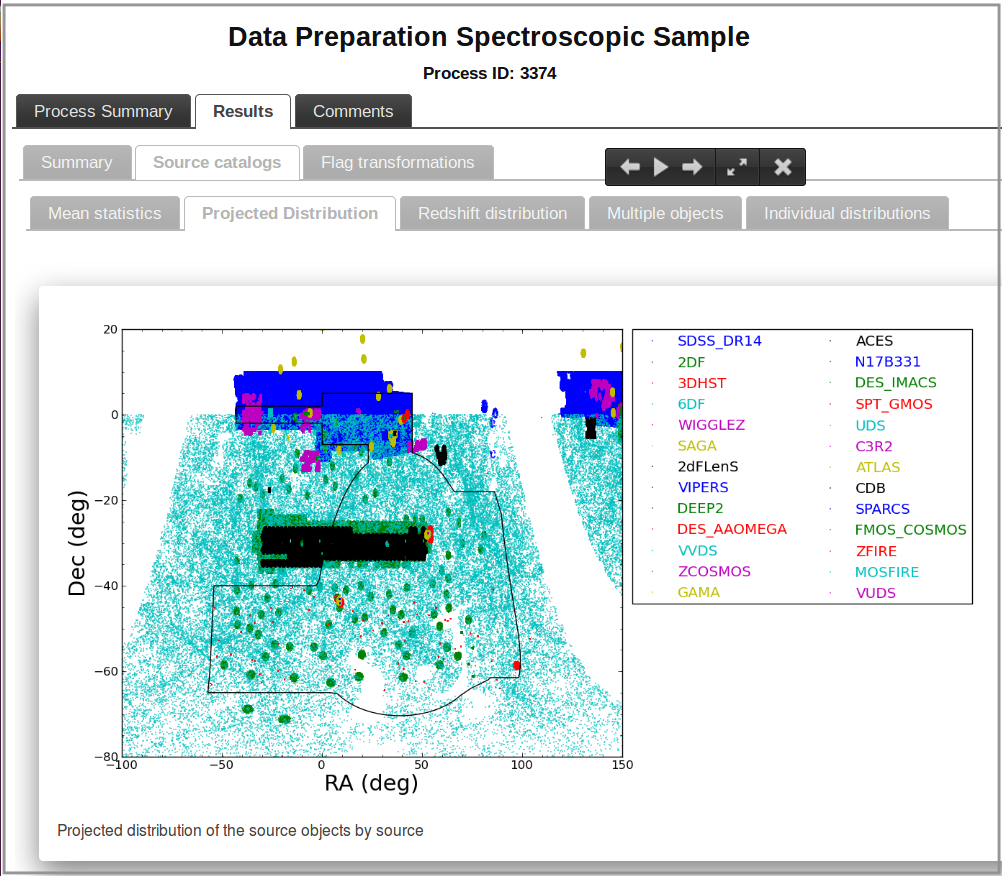}
    \end{center}
    \caption{Product log of \textit{\textbf{Spectroscopic Sample}} pipeline. Screenshot capturing one of the plots available on the ``Results'' tab of the product log. The black line represents the DES footprint. The surveys in the legend are ordered by the number of spectroscopic sources included in the sample after resolving the duplicates.}
    \label{fig:spec_prodlog}
    \end{figure*}

Furthermore, the supplemental video V1\footnote{\url{https://youtu.be/1mu-PqOvK88?list=PLGFEWqwqBauBIYa8H6KnZ4d-5ytM59vG2}} shows an example of a guided run of the \textit{\textbf{Spectroscopic Sample}} pipeline and a quick exploration of its results.

    \subsection{Training Set Maker}
    \label{subsec:ts_maker}

After creation of the spectroscopic sample, the next step is to match the photometric data to the spectroscopic catalog. This is done by the \textit{\textbf{Training Set Maker}} pipeline, designed to build training (and validation) samples by matching a photometric sample (among the datasets available in the Portal) with a spectroscopic sample, which comes from the \textit{\textbf{Spectroscopic Sample}} pipeline, as shown in Figure~\ref{fig:ts_maker_input}. 

In this stage the user can also include outputs of the \textit{\textbf{Star/Galaxy}} \textit{\textbf{Separation}} pipeline, identifying and removing point sour\-ces, avoiding the mismatch between spec-$z$s from galaxies mer\-ged to photometry from stars and removing QSOs. Also, optionally, it is possible to apply corrections in the observed magnitudes, like zero--point based on stellar locus regression calibrations \citep{Hig09} or galactic extinction. 

\begin{figure*}
    \begin{center}
    \includegraphics[width=1.7\columnwidth]{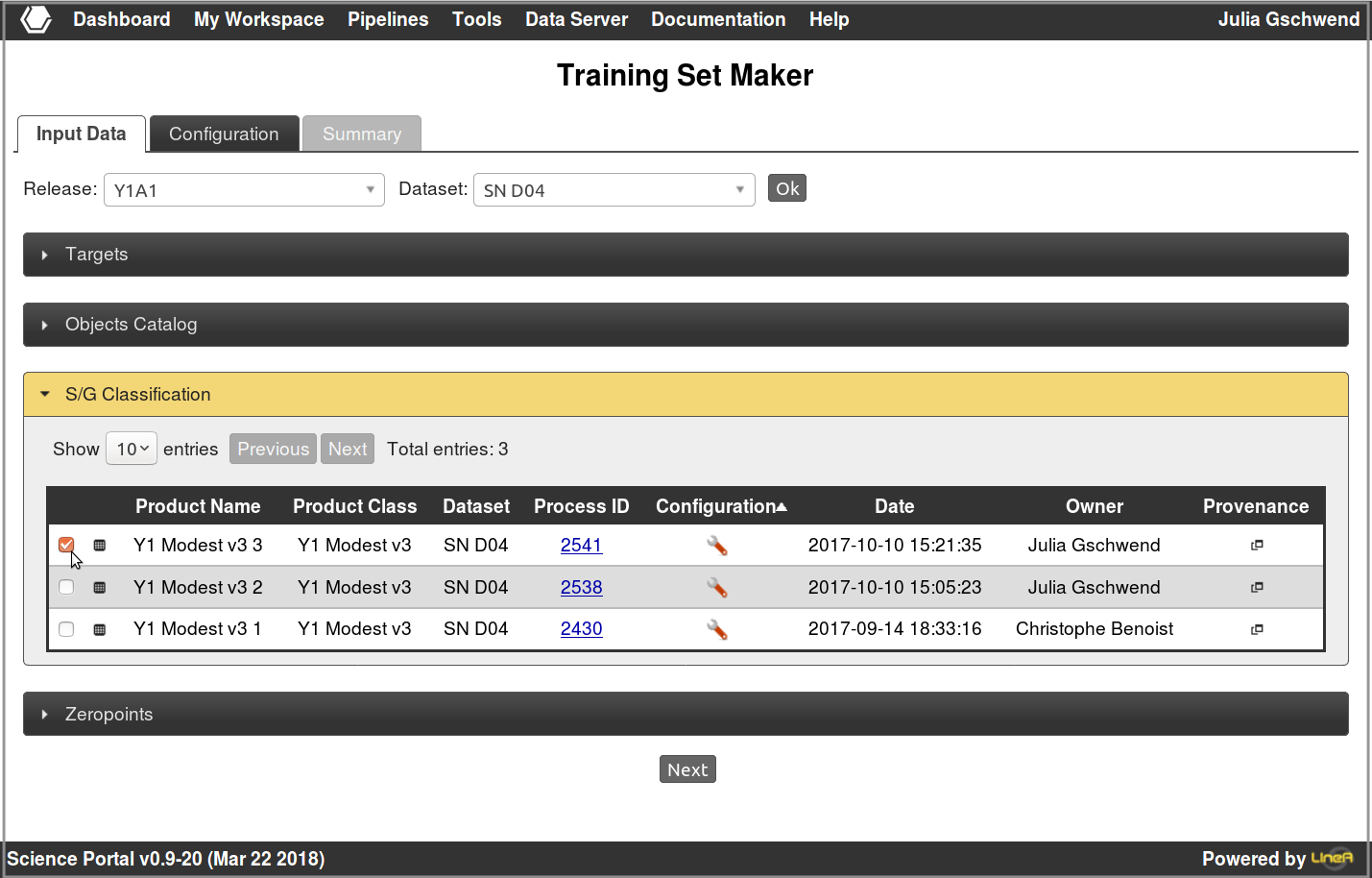}
    \end{center}
    \caption{Input menu of the \textit{\textbf{Training Set Maker}} pipeline. Each one of the four pull down menus show the available products for each class (mandatory or not) required by this pipeline: \textit{Targets} -- the spectroscopic sample defined in Section~\ref{subsec:pz_train}, \textit{Object Catalog} -- the coadd tables coming from DES, \textit{S/G Separation} -- a star/galaxy  classification table, provided by another pipeline, not addressed in this paper, and \textit{Zeropoints} -- optional additional photometric calibrations. Each row with a check box refers to a product generated by a previous pipeline, registered in the database. The process number is a link that redirects to the process' product log, which helps the user on the choice.} 
    \label{fig:ts_maker_input}
\end{figure*}

In the configuration menu (Figure\ref{fig:ts_maker_config}), the user selects the parameters to make quality choices and a search radius used for matching. Similarly to the previous pipeline, the matching is also done based on the angular separation between the objects at the database level using Q3C, but here with the spectroscopic and photometric catalogs. We also selected the radius to 1.0 arcsec as a default configuration. If two or more objects from the photometric sample are within the search radius, the nearest object to the spectroscopic one is selected. 

\begin{figure*}
    \begin{center}
    \includegraphics[width=1.7\columnwidth]{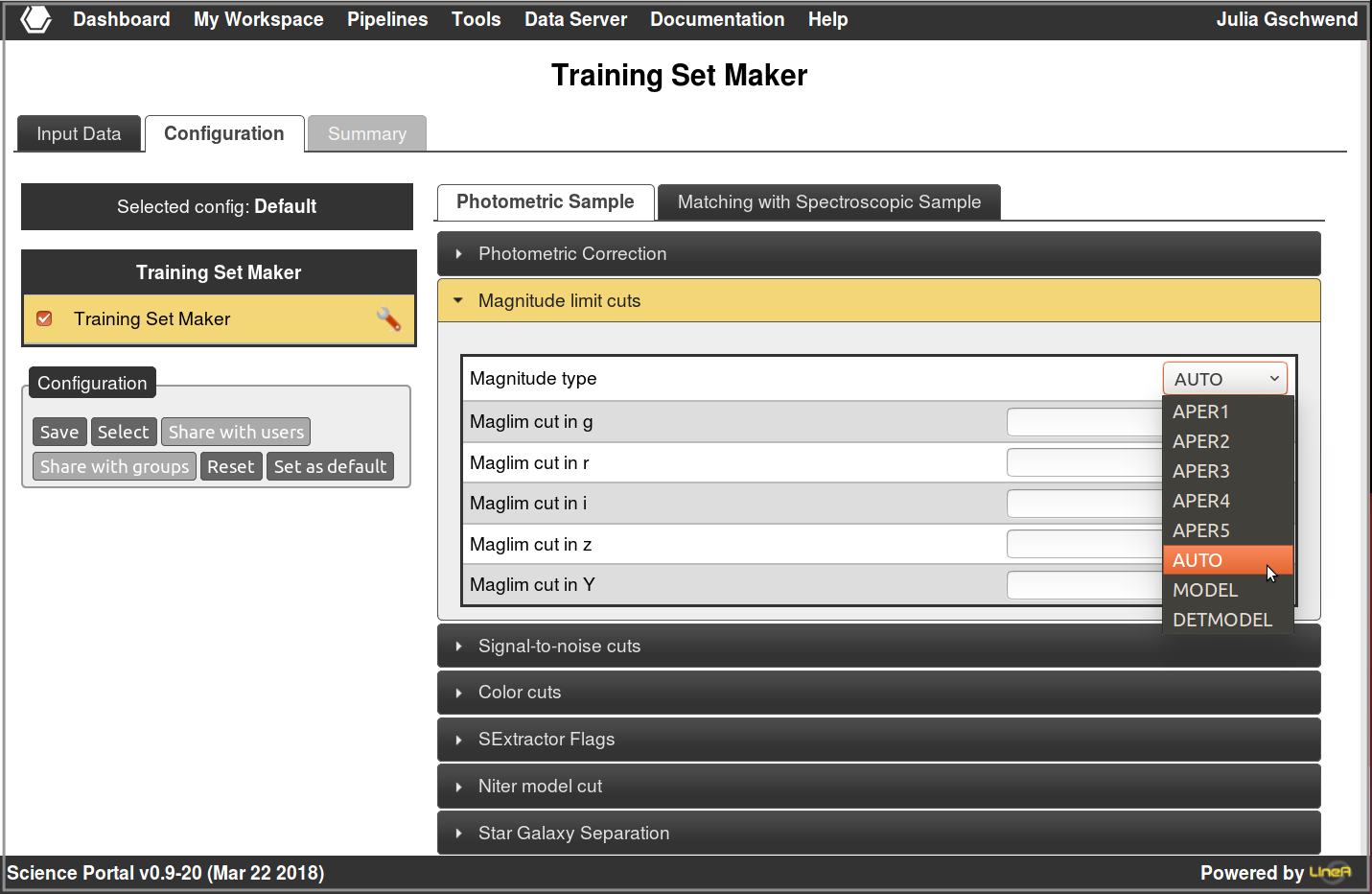}
    \end{center}
    \caption{Configuration menu of the \textit{\textbf{Training Set Maker}} pipeline. On this screen, the user is asked to make decisions about the characteristics of the matched sample, called training set, that is being constructed. In this example, the type of magnitude used to apply quality cuts. }  
    \label{fig:ts_maker_config}
\end{figure*}

The result of this pipeline is a table registered in the database under the class ``training\_set'', containing the columns from the spectroscopic sample plus some columns from photometric data (DES IDs, magnitudes, and errors). On the product log, one can access the query automatically generated by the pipeline (as illustrated in Figure~\ref{fig:ts_maker_query}) and pieces of information about the matched sample, so--called training set, organized in some tabs. In this example run, we selected the spectroscopic sample defined in the previous example (Figure~\ref{fig:spec_prodlog}) and the photometric sample from DES Y1 wide survey (details in \ref{app_sub:phot_sample}).  

We present an example of this operation in the supplemental video V2\footnote{\url{https://youtu.be/2nA1PFGCnEM?list=PLGFEWqwqBauBIYa8H6KnZ4d-5ytM59vG2} } showing a live run of the pipeline \textit{\textbf{Training Set Maker}}, using the same inputs and configurations as shown above.  

\begin{figure*}
    \begin{center}
    \includegraphics[width=1.7\columnwidth]{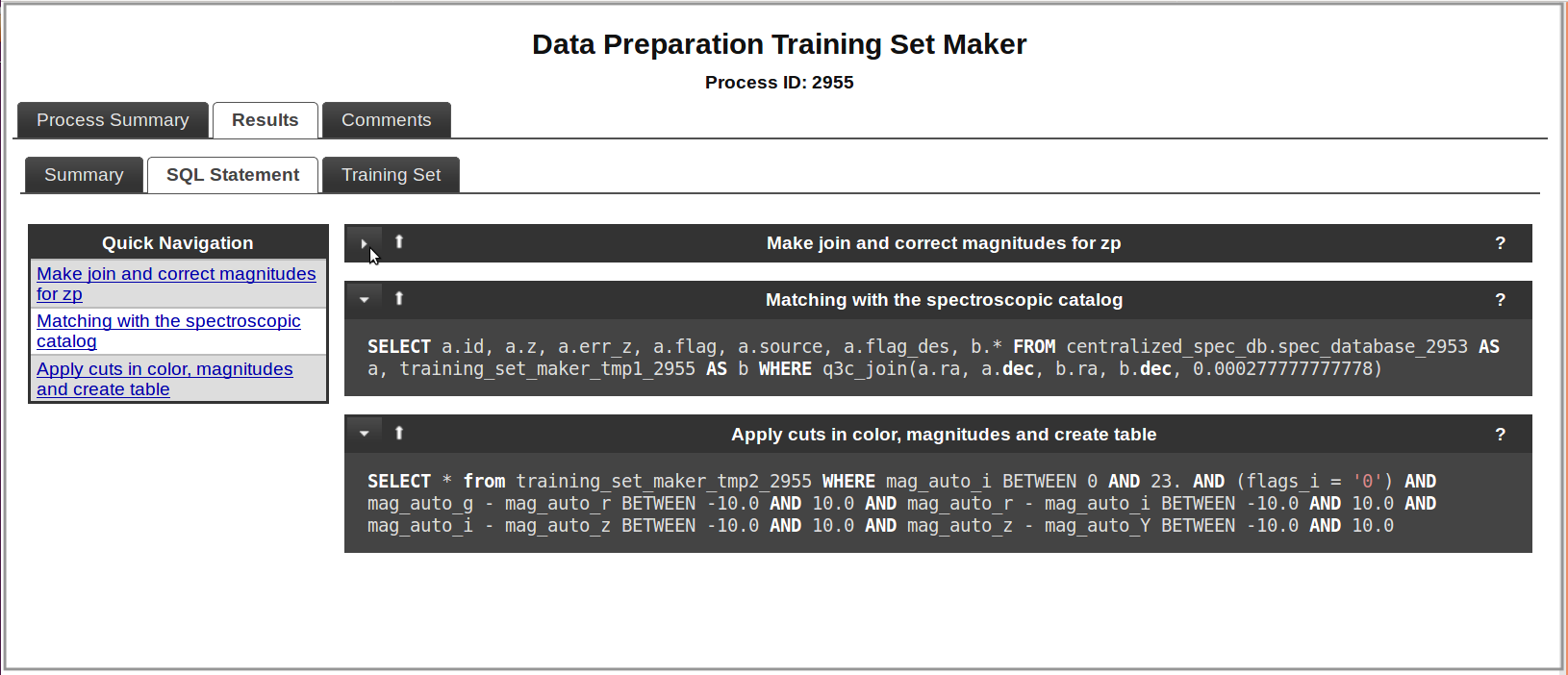}
    \end{center}
    \caption{Product log of \textit{\textbf{Training Set Maker}} pipeline. This screen summarizes the results of a pipeline run. The SQL query used by the pipeline is always shown. In the second tab, there are tables and plots to show the characteristics of the training set just created.}
    \label{fig:ts_maker_query}
\end{figure*}

    \subsection{Photo-z Training and Validation} 
    \label{subsec:pz_train}

After the matched sample is created and registered in the portal's database, the next step before calculating photo-$z$s for the whole photometric dataset is to train the empirical algorithms and, optionally, calibrate the template--fitting ones. For several science applications, it is necessary to know the quality of the photo-$z$ estimations requiring a validation step also done with a sample with known spectroscopic redshifts. 

Although it is not the pipeline with the largest number of components, \textit{\textbf{Photo-$z$ Training}} is the most complex among pipe\-lines related to photo-$z$, because it performs several different tasks, as detailed below. It is composed of three components, as illustrated in  Figure~\ref{fig:photoz_flowchart}. The first is \comp{Subsets} \comp{Separation}, which splits the matched sample into two subsets and performs a comprehensive characterization of them with plots and statistics.  The second is the \comp{Photo-z} \comp{Training}, which conducts the training procedure using the first subset, for several algorithms simultaneously. The third is the \comp{Photo-z} \comp{Validation} which uses the second subset to compute the photo-$z$s and compare the results with the spec-$z$s, as well as it shows metrics and plots for quality assessment. 

The primary input for the \textit{\textbf{Photo-z Training}} pipeline is the matched sample built by the previous pipeline, named as ``Training Set'' on the input menu shown in Figure\ref{fig:pz_train_input}. Also, we choose a photometric sample of reference (``Objects catalog'', on the input menu), the same coadd tables mentioned in the previous pipeline.  But in this case, this one is used only to compare photometric properties, to check whether the training and validations subsets are representative of the photometric set or not. 

In the first implementation, we created a unique pipeline to do both steps, the training and validation (The pipeline \textit{\textbf{Photo-z Training}}). Thus, a first component is necessary to separate the data in training and validation subsamples (to avoid the biases introduced by validating with the same galaxies used for the training). Afterward, we created another pipeline to do the validation step separately (details below), but we kept the first option available, optionally. 

The \comp{Subsets} \comp{Separation} component splits the matched catalog randomly, where the fraction of the data delivered to the training subset (and consequently the remaining portion for validation) is a free parameter in the component's configuration. Also, the user can define the sample selection criteria, choosing the acceptable intervals of magnitude, redshifts, colors and magnitude signal--to--noise ratio, as illustrated in Figure~\ref{fig:pz_train_subset_config}. 

Ideally, the training and validation samples should have the same properties as the photometric sample of interest. However, this is difficult to meet when spectroscopic data come from surveys with different depths, redshift intervals, and targeting strategies. In cases like this, it is a common practice to evaluate the performance of the photo-$z$ in a {\it weighted} sample, representing the color and magnitude distributions of the photometric sample.

\begin{figure*}
    \begin{center}
    \includegraphics[width=1.7\columnwidth]{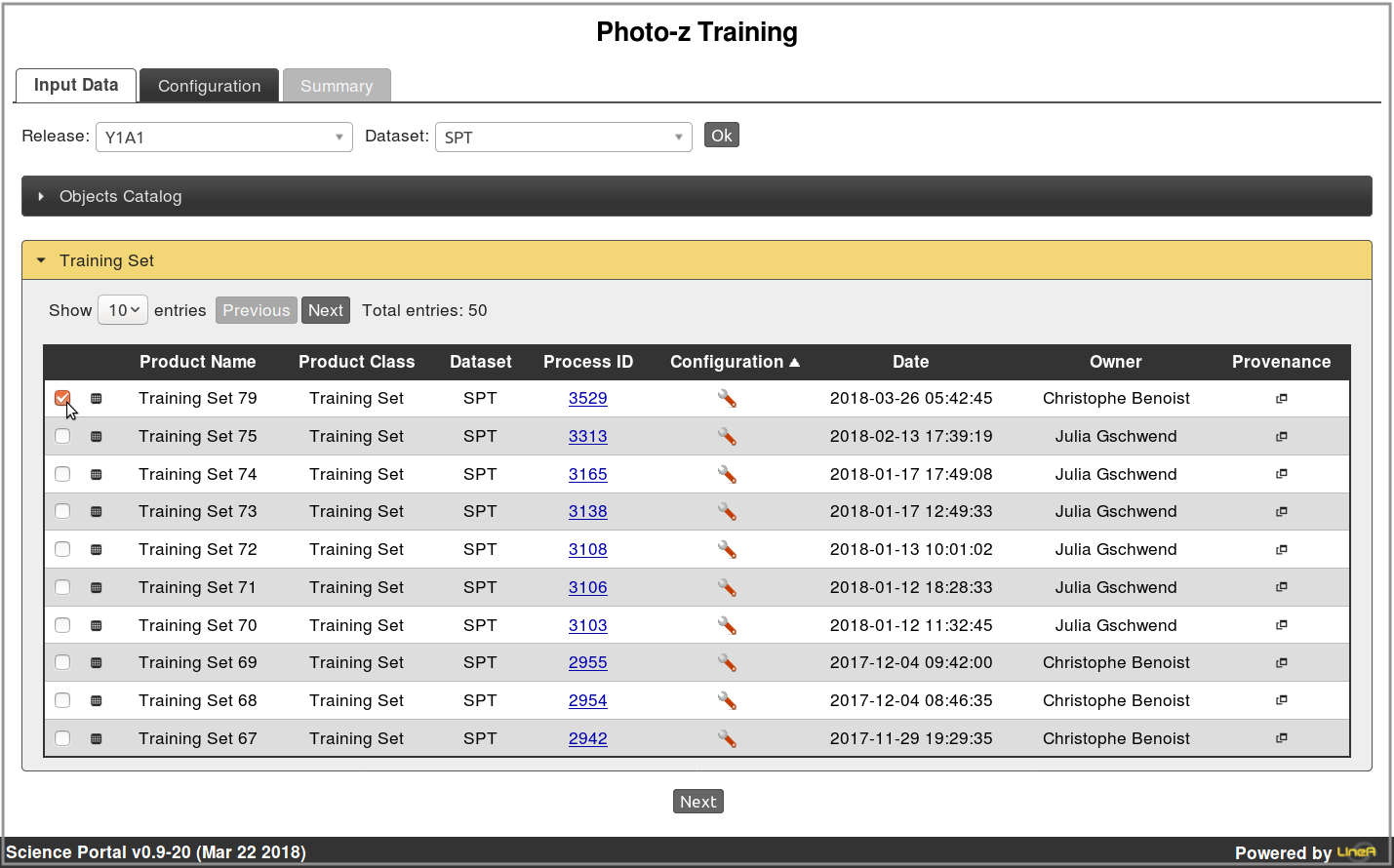}
    \end{center}
    \caption{Input menu of the \textit{\textbf{Photo-z Training}} pipeline. On this screen, the user is asked to chose one matched sample (recognized by its class ``Training Set''), and a photometric sample of reference (``Objects Catalog''). Similarly to any other pipeline, each row with a check box refers to a product previously generated by another pipeline, and registered in the database. The process number is a link that redirects to the process' product log, which helps the users on their choice.}
    \label{fig:pz_train_input}
\end{figure*}

\begin{figure*}
    \begin{center}
    \includegraphics[width=1.7\columnwidth]{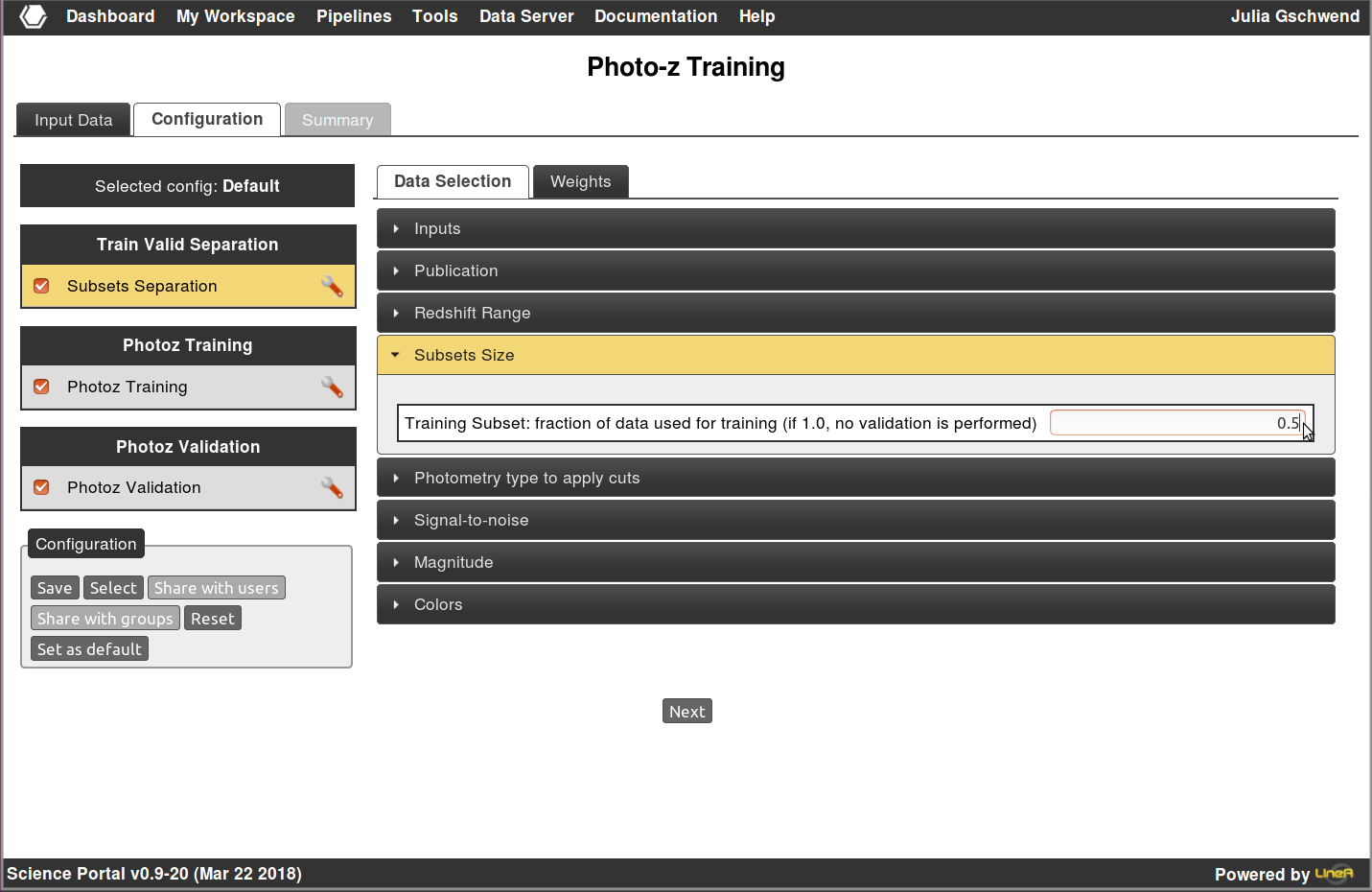}
    \end{center}
    \caption{Screenshot of the \textit{\textbf{Configuration}} tab of the component Subsets Separation, that belongs to the \textit{\textbf{Photo-z  Training}} pipeline. The default value for the fraction of data given to training and validation is 0.5 for each subset. If the fraction is chosen to be 1.0, the whole sample is employed for training, and the validation step is skipped.} 
    \label{fig:pz_train_subset_config}
\end{figure*}

In the Portal, it is optional to weight the training and validation sets, using the algorithm presented in \cite{Lim08}. If so, we assign to each galaxy its relative importance in representing the photometric sample, regarding the multi--space of colors and magnitudes. The user builds the \textit{weighted} sample by repeating galaxies multiple times in the proportion of their weights, with their magnitudes spread according to their errors (assumed Gaussian) to avoid generating identical cloned galaxies. Applying this algorithm, we obtain a weighted sample that presents distributions of colors and magnitudes very similar to those of the photometric sample, as shown in Figure~\ref{fig:hist_i_gr_specz}. In the example, the excess of red objects in the training set is diminished as the result of weighting.

\begin{figure*}
    \begin{center}
    \includegraphics[width=1.50\columnwidth]{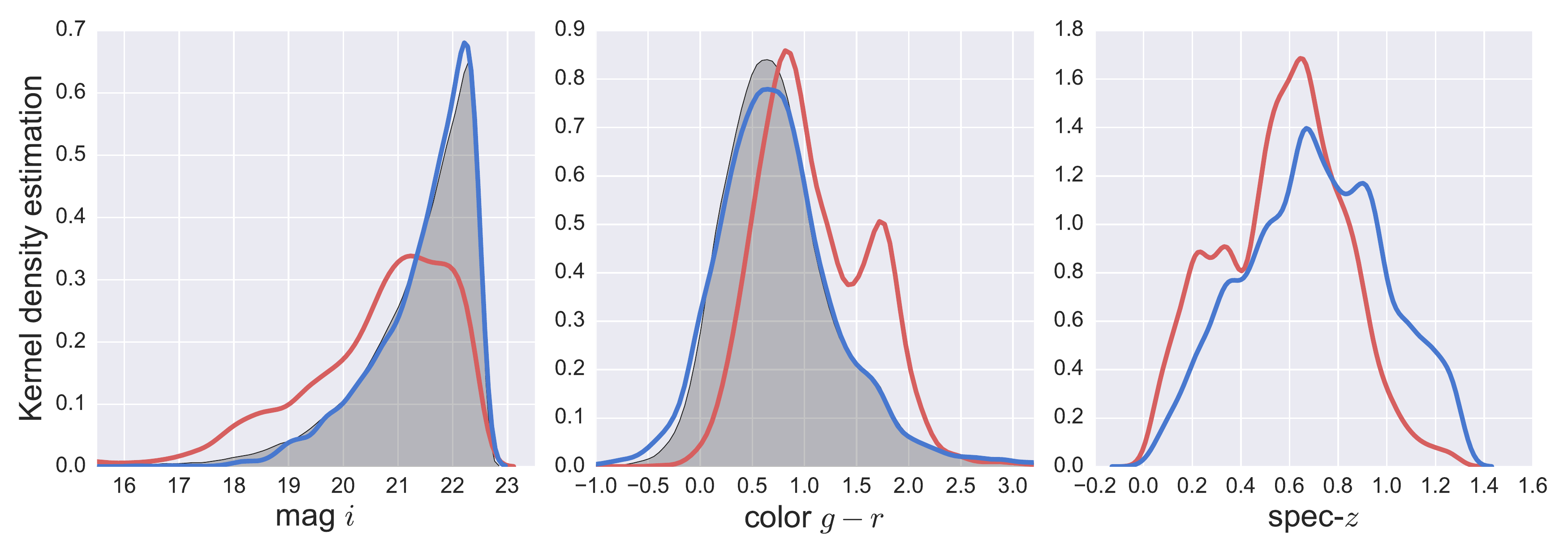} 
    \caption{Magnitude ($i$--band), color ($g$-$r$), and spec-$z$ distributions for the photometric sample (Y1A1, in gray), and for the training set, before (in red) and after (in blue) weighting. The validation set, not shown here, has the same properties as the training set.} 
    \label{fig:hist_i_gr_specz}
    \end{center}
\end{figure*}

    \subsubsection*{Photo-z Training}

In recent years the number of photo-$z$ algorithms has increased enormously. The Portal is an interesting environment to compare different methods, since they can be applied to datasets under similar conditions. So far, the following codes are implemented in the Portal: \textsc{Annz} \citep{Col04}, \textsc{Annz2} \citep{Sad16}, \textsc{ArborZ} \citep{Ger10}, \textsc{BPZ} \citep{Ben00}, \textsc{DNF} \citep{deV15}, \textsc{LePhare} \citep{Arn02, Ilb06}, \textsc{Pofz} \citep{Cun09},  \textsc{Sky-Net} \citep{Gra14}, and \textsc{TPZ} \citep{Car13, Car14}. We refer to \citet{Hil10} and \citet{Car14} for a review of the particularities and comparison of their performances.

Empirical methods are the basis for the majority of the algorithms, except for \textsc{BPZ} and \textsc{LePhare}, two template fitting codes, for which a training sample can be used to improve photo-$z$ quality through systematic shifts in the theoretical magnitudes from the spectral energy distribution (SED) templates. Hence, all of them are implemented in \textit{\textbf{Photo-z Training}} pipeline. Nevertheless, for the template--fitting ones, the ``training'' step is not mandatory. 

Each photo-$z$ algorithm has its configuration parameters. The user interface provides a configuration menu with a default configuration, but the user can change these values as shown in Figure~\ref{fig:pz_train_dnf_config}. 

The product of this training procedure is the so--called \textit{training file}. Its format and content depend strongly on the photo-$z$ algorithm used. For instance, \textsc{TPZ}'s training files are stored in NumPy\footnote{\url{http://www.numpy.org/}} format files, containing the decision trees used in photo-$z$ estimation.  \textsc{LePhare}'s training files are just a list of floating point numbers representing the systematic shifts applied to the theoretical magnitudes (those obtained from the SED templates), stored in a simple text file. Besides the training files, the component \comp{Photo-z} \comp{Training} also registers the code configurations used, so it is also applied by the pipeline \textit{\textbf{Photo-z Compute}}, where the photo-$z$s are estimated for the DES datasets.

\begin{figure*}
    \begin{center}
    \includegraphics[width=1.6\columnwidth]{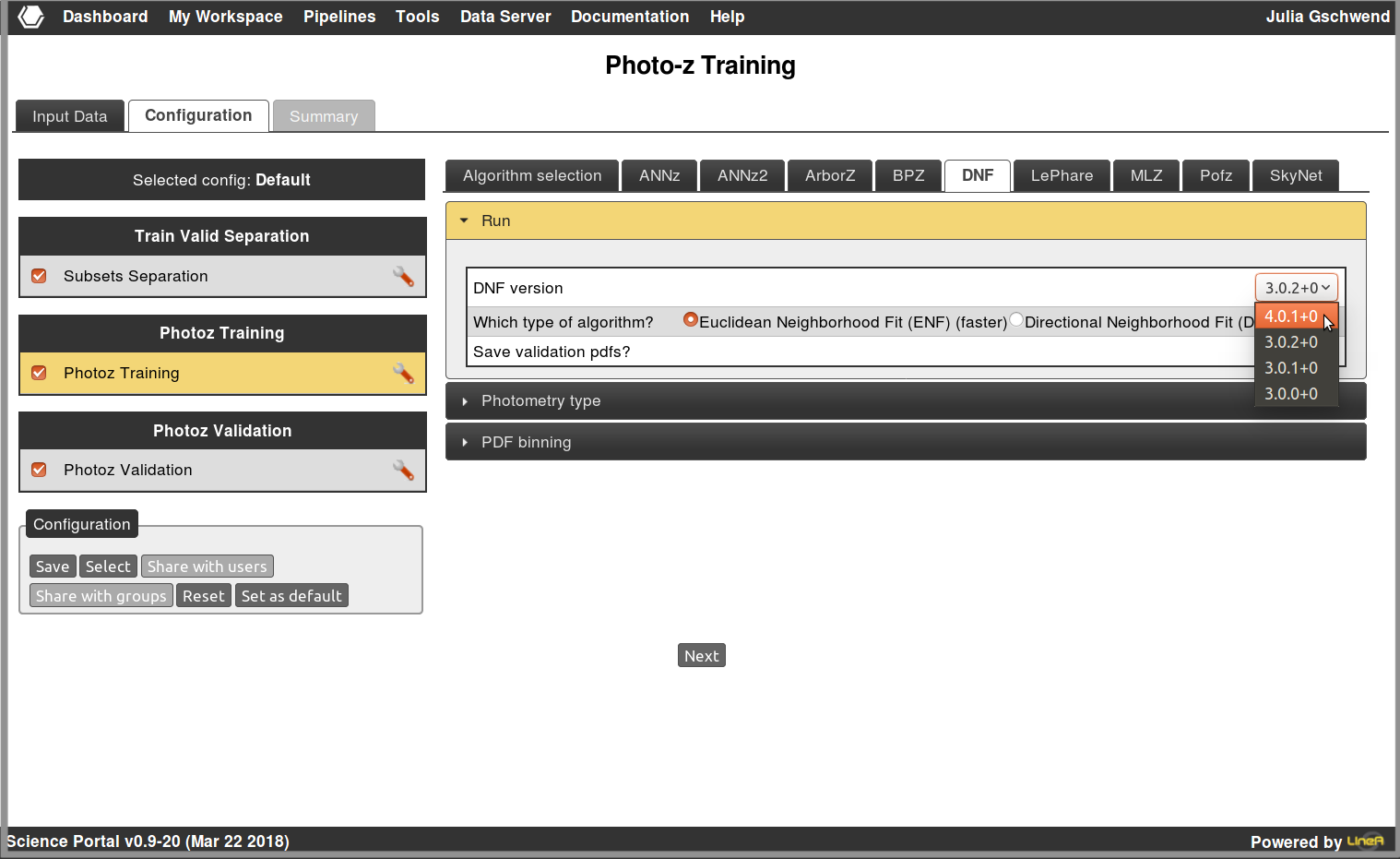}
    \end{center}
    \caption{Screenshot of the configuration tab of the \textit{\textbf{Photo-z Training}} pipeline. On the left side, each check box refers to one component of this pipeline. The small tool symbol beside leads to the menu on the right side of the page, where several tabs organize the configuration parameters for the different algorithms available. In this screenshot we show those from the code BPZ as an example.}
    \label{fig:pz_train_dnf_config}
\end{figure*}

The main advantage of performing this step independent from the actual photo-$z$ estimation is that training files from one training procedure can be used in the photo-$z$ calculation several times, for different photometric datasets. On the other hand, one can make training and validation several times, until gets a satisfactory result and then apply it to the photo-$z$ central estimate.

    \subsubsection*{Photo-z Validation}

The last component of \textit{\textbf{Photo-z Training}} pipeline is the \comp{Photo-z} \comp{Validation}. It is responsible for checking the quality of the photo-$z$s computed in a validation sample, as an estimate of the quality of the photo-$z$ to be estimated for the large photometric datasets.

To meet science--driven requirements, sometimes one needs to perform training and validation in samples which are independent of each other. Hence, we created a new pipeline (keeping the first one active) called \textit{\textbf{Photo-z Validation}} to perform only the validation step, using the result of training from a previous run of the pipeline \textit{\textbf{Photo-z Training}}, but with the possibility to receive a completely different matched sample as input data. This pipeline uses the same component \comp{Photo-z} \comp{Validation} as the \textit{\textbf{Photo-z Training}} pipeline, therefore the methodology is the same. The coincidence of pipeline and components' names might lead the reader to a confusion. We clarify the sequence of tasks performed by the components grouped by the pipelines in Figure~\ref{fig:photoz_flowchart}.

In summary, there are two possible ways to validate photo-$z$s in the Portal: (i) splitting the matched spec--photo sample (so--called training set) into two subsets and perform the validation at the last component of \textit{\textbf{Photo-z Training}} pipeline; (ii) training with 100\% of the training set, and do the validation separately, in another pipeline, with an independent validation set. Both ways follow the same methodology. The only difference is the definition of the inputs.

\begin{figure*}
    \begin{center}
    \includegraphics[width=1.7\columnwidth]{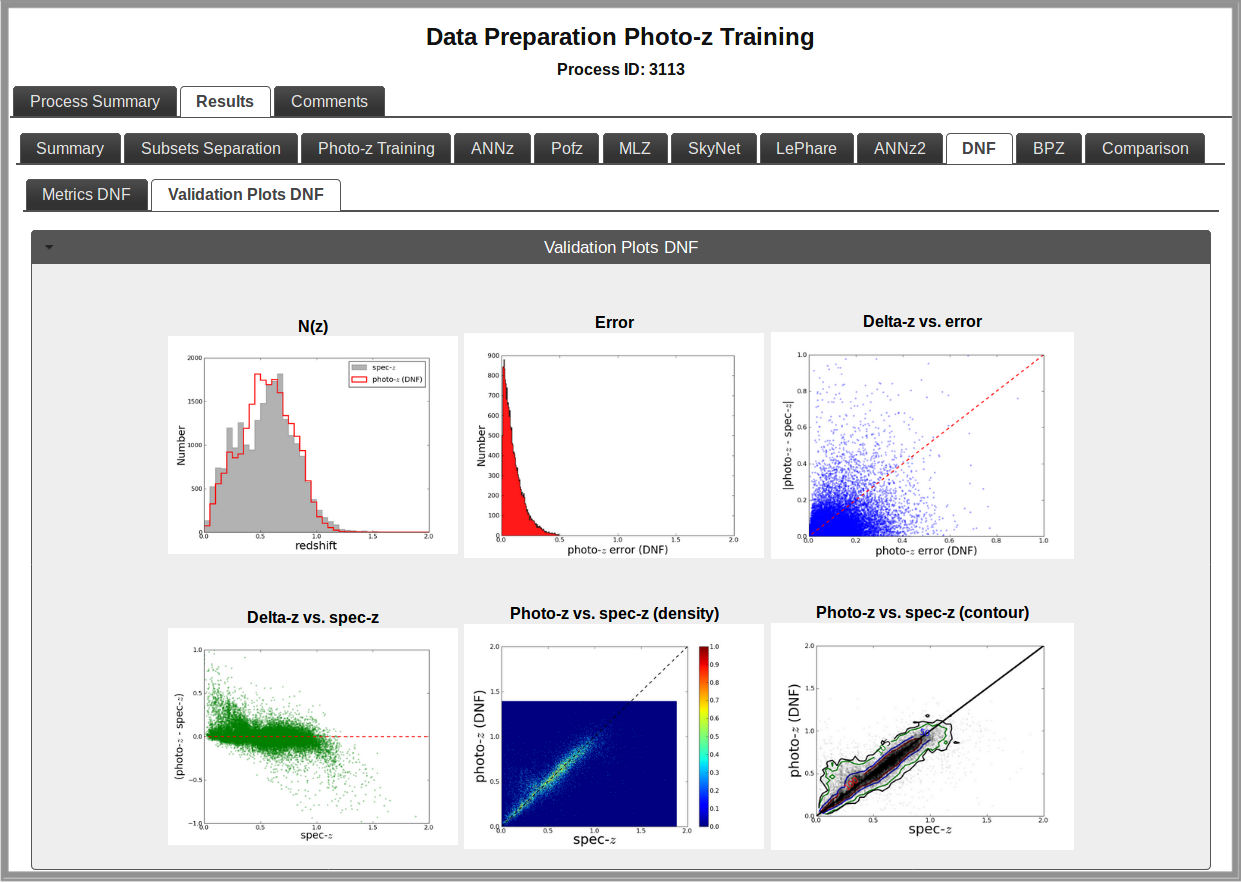}
    \caption{Results from the \comp{Photo-z} \comp{Validation} component, organized in tabs by photo-$z$ algorithms. In this example, we present results by \textsc{DNF}. The three plots on the top are, from left to right, the histograms of redshift and error distributions, and the scatter plot of photo-$z$ error versus the difference between photo-$z$ and spec-$z$ for each object in the validation sample. The plots on the bottom are the scatter plot of photo-$z$ versus the difference between photo-$z$ and spec-$z$, and the density and contour plots of photo-$z$ versus spec-$z$.  }  
    \label{fig:pz_valid_plots}
    \end{center}
\end{figure*}

The validation results consist of photo-$z$ metrics (to quantify bias, dispersion, etc.) and quality assessment plots for visual inspection. The definition of the metrics used can be found in \citet{San14}. Uncertainties in the metric values are estimated using the Bootstrap re--sampling technique \citep{Bra93} based on 100 realizations, as done in such work. Some of these metrics have a limit of acceptance, defined by the collaboration as a scientific requirement for dark energy studies. So this component also works as a ``vetting point'' for the photo-$z$ estimates. If the photo-$z$ quality is considered unacceptable, the user should repeat the previous steps varying the data used and the configuration parameters. 

An example of product log is presented in Figure~\ref{fig:pz_valid_plots}, showing the results obtained using \textsc{DNF}. This figure shows how this pipeline can be used to compare performances of different algorithms like the ones done by \citet{Hil10}, and \citet{San14}.

The user can navigate through tabs to access the results from different codes. In particular, there is an additional tab where the results are consolidated and presented together to ease the comparison. 

For more detailed navigation through the various configuration parameters and results reported on the product log, please watch an example of usage of the \textit{\textbf{Photo-z Training}} pipeline in the supplemental video V3\footnote{\url{https://youtu.be/ZOJ0hGWlvag?list=PLGFEWqwqBauBIYa8H6KnZ4d-5ytM59vG2}}.

    \subsection{Photo-z Compute} 
    \label{subsec:pz_compute}

The actual photo-$z$ calculation in the Portal is done by the \textit{\textbf{Photo-$z$ Compute}} pipeline. It estimates photo-$z$s for DES objects present in the photometric catalogs, regardless of the object's nature (e.g., star or galaxy), using the training file(s) produced by \textit{\textbf{Photo-z Training}}. Once the photo-$z$s are calculated, they can be used in the creation of science--ready catalogs considering, e.g., different star/galaxy classifiers, color selections, and magnitude limits. It is also possible to download the photo-$z$ resulting tables, or to deliver them to the collaboration through the export tool, connected to DESDM database. Since the same object can be tagged as galaxy by one classifier or star by another, it is essential to have photo-$z$s available for all objects. Therefore, at this stage, the distribution of redshifts, $N(z)$, obtained with the \textit{\textbf{Photo-$z$ Compute}} pipeline is not representative of the galaxy distribution yet. Only later, when the final catalogs are produced after pruning, the $N(z)$ can be used for scientific analyses.

\begin{figure*}
    \begin{center}
    \includegraphics[width=1.7\columnwidth]{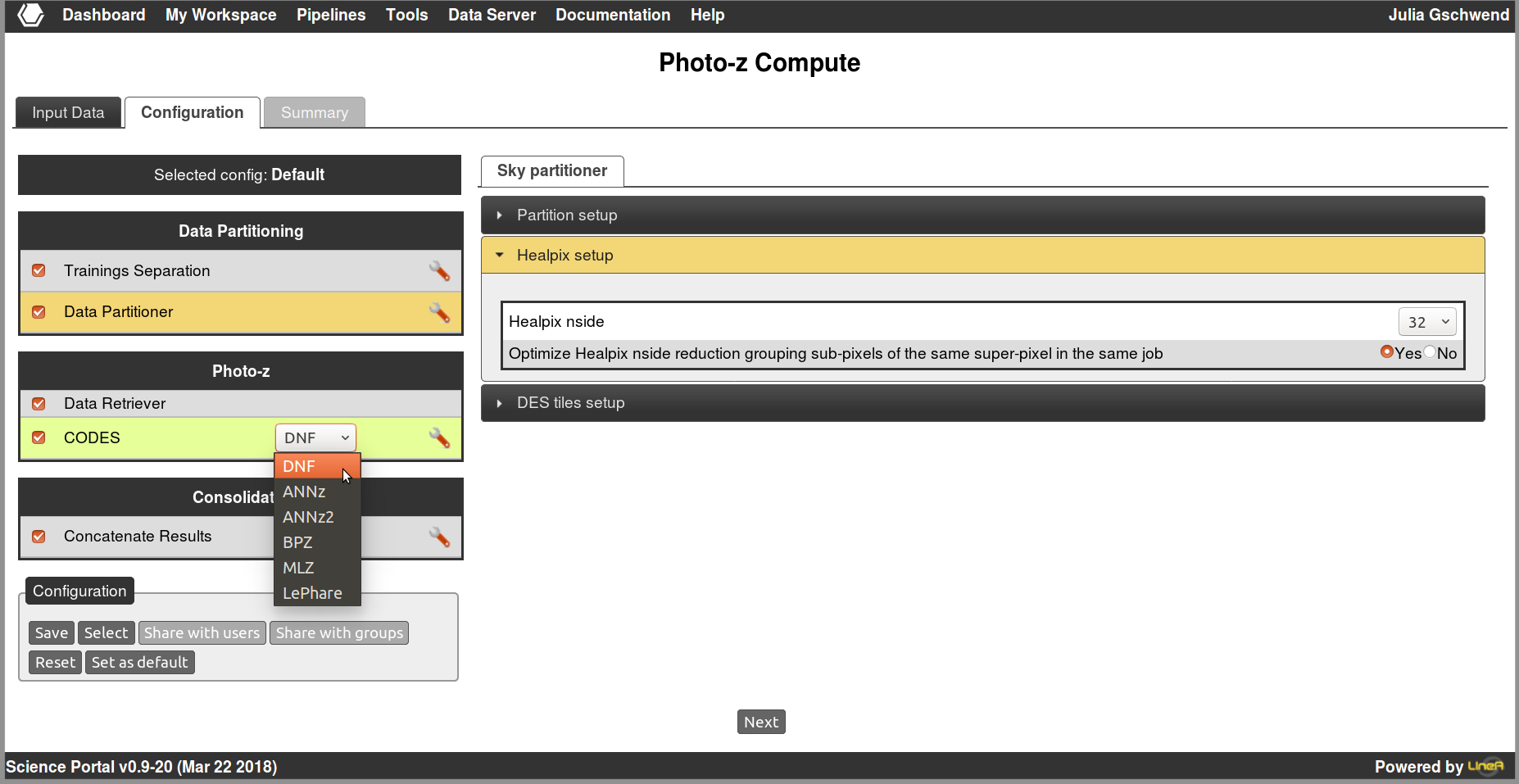}
    \end{center}
    \caption{Screenshot of the configuration tab of the \textit{\textbf{Photo-z Compute}} pipeline. On the left, the five components mentioned in the text. For this example, we chose DNF as the photo-$z$ algorithm to be applied by the fourth component. On the right, the choice of the size of each data partition, based on HEALPix pixels.} 
    \label{fig:pz_comp_config}
\end{figure*}

The \textit{\textbf{Photo-$z$ Compute}} is composed of five components. The dashed line in Figure~\ref{fig:photoz_flowchart} highlights the part of the workflow that runs in parallel processing. The first component, \comp{Photo-z} 
 \comp{Separate} handles with the training files inherited from the previous process. The second component, \comp{Sky} \comp{Partitioner},  defines the data partition by dividing the area in the Sky covered by the input dataset, based on the user choice of partition unit (as detailed below). It distributes the information about the data partitions into the nodes where the photo-$z$ codes run in parallel.

The next two components (enclosed by the dashed line in Figure~\ref{fig:photoz_flowchart}) run in parallel, each item of the partition defined by the previous one. The component \comp{Partition} \comp{Retriever} loads the contents of data partition in each computing node. To deal with large photometric samples efficiently, the data access uses HDFS, which previously distributed chunks of data in the cluster nodes, minimizing time spent with data transfer. Therefore, the  \comp{Partition} \comp{Retriever} reads from the cluster nodes, as discussed in Section~\ref{sec:portal}.

It is only in the fourth component that the photo-$z$s are in fact estimated. This step is the most computationally intensive of all the tasks related to photo-$z$ estimation. There is one component available for each algorithm. All of them contain a python wrapper that prepares the input data, runs the code, standardizes the outputs, and delivers it to the consolidator. Thus, the workflow calls only the one that corresponds to code chosen in the configuration screen (see the menu on the left side displayed in Figure \ref{fig:pz_comp_config}).

The consolidated jobs and their resulting photo-$z$ table are ingested in the database by the last component, called \comp{Join} \comp{Photo-z} \comp{Compute}. The product of \textit{\textbf{Photo-z Compute}} is one of the leading ingredients to compose a science--ready catalog for extragalactic sciences, as discussed in \citet{Fau18}.

Since the Portal provides flexibility on the parallelization strategy, we performed a series of tests to illustrate the use of our computer cluster. In the following paragraphs, we compare the computing of photo-$z$s by varying the size of the data chunk, and consequently, the number of data partitions. For simplicity, we use the original data division from DES, based on tiles (see discussion ahead). The methodology details of the parallelization are discussed in Section~\ref{sec:portal}. 

We use the most extensive dataset of Y1A1 data release (SPT, details in \ref{app_sub:phot_sample}) to make stress tests and test the cluster capacity. In this case, we choose to use the algorithm DNF, which is one of the fastest codes available in the Portal, according to previous tests not addressed in this work. 

The most straightforward way to vary the parallelization strategy is to modify the number of tiles to be processed by each job submitted to the cluster management system. Table~\ref{table:parallel_results_tiles} and Figure~\ref{fig:pz_compute_by_tiles_SPT} summarize the results of the tests. In this table, the first column shows the number of tiles processed per job in parallel. The second column shows the total number of jobs, which is approximately the total number of tiles in SPT (3,373) divided by the first column. The third column shows the time spent in the serial parts of the processes (organizing training files, defining the partitions, and concatenating the results at the end). The fourth column shows the parallelized part of the processes (the data retrieving and the actual photo-$z$ estimation, the components surrounded by dashed line in Figure~\ref{fig:photoz_flowchart}). The fifth column shows the total duration of each process. 

All processes started with the same inputs and code configuration and delivered the same results. The only difference was in the definition of data partitions, which was seen to have a significant impact on the process duration.  Hereafter, we refer to  ``infrastructure time'' as the difference between the total time of a process and the time spent on actual code running. This quantity is difficult to measure when running in parallel. It is often the case where a group of jobs is submitted to the computer cluster, virtually simultaneously, but they do not finish at the same time, even though all the nodes have the same hardware characteristics and the sizes of the data chunks are virtually homogeneous.

Some of the possible reasons for the different delays are: i) reading data from the same node versus reading data from a neighbor node; ii) long queues of data partitions waiting for their jobs to start; iii) bottleneck for writing in the 'reduce' part of the workflow (where some jobs still waiting in a queue to register the results, when others are already writing). 

Another contribution to the infrastructure time might be the time spent by HTCondor to manage the jobs (start, finish, writing logs, creating temporary directories, and distributing jobs in the cluster).
Although we can raise several possible reasons for the time lost in processes and the differences in time delays between processes with different partition sizes, we can not measure precisely relative contribution from each one of these sources of delay. Therefore, the infrastructure time is the cumulative time loss due to a combination of reasons. 


\color{black}

We recall that there is an option to reserve the entire node (24 cores) for a single job, when the process deals with internal parallelization, as mentioned in Section~\ref{sec:portal}. This is not the case here. For this test, this option of node reservation is disabled, so the jobs are distributed all over the cluster, regardless of the nodes to which the cores belong. Therefore, the maximum number of jobs running simultaneously is 912 (38 nodes times 24 cores). This number possibly explains why $N=4$ is the most effective strategy for this first test. If the infrastructure time was null and the execution time of the primary algorithm was proportional to the size of the data chunk then it would be virtually equivalent to run 1 or 4 tiles per core, i.e., 1 round of 843 jobs running four tiles each, or 4 rounds of 912 (actually 3 rounds of 912 plus one of 637) jobs running one tile each. Since the infrastructure time is not zero, and it is cumulative, four tiles per node are better than 1, because it occupies almost the whole cluster with jobs, without leaving any other job waiting on a queue. The small queue of four files to be processed within a job seems to work more efficiently than a large queue of individual files distributed to the cluster. 

As expected, for $N>4$, the larger the $N$, the longer the processing time. This is true because large $N$ reduces the number of CPU cores used, wasting the capacity of the cluster. {In summary, according to our tests, the optimal number of tiles distributed per job depends on the dataset size as the following: If the number of tiles of a dataset is less than 912 (maximum number of jobs running in parallel) than the optimal choice is to distribute one tile per job. If the total number of tiles is larger than 912, the optimal number of tiles per job is the integer number closest to the result of the total number of tiles divided by 912.  In the example shown in Figure 22, the total number of tiles in the SPT dataset, 3373, divided by 912 is $\sim$3.7. Therefore, the fastest run using this particular dataset was the one with four tiles per job, as observed in Figure~\ref{fig:pz_compute_by_tiles_SPT} and Table~\ref{table:parallel_results_tiles}.} Interestingly, the data partitioner and the consolidator performances are very stable, independently of the number of partitions they handle.

\begin{table}
    \begin{center}
    \caption{Execution time of pipeline \textit{\textbf{Photo-z Compute}} - dataset SPT, data partition based on DES tiles. } 
    \label{table:parallel_results_tiles}
    \begin{tabular}{*{5}{c}}

\hline
\# Tiles/job & \# Jobs & Serial$^{\dagger}$ & Parallel$^{\dagger}$ & Total$^{\dagger}$  \\ 
\hline
1  & 3,373 & 00:49 & 02:30  & 03:19  \\
4  &   843 & 00:47 & 01:19  & 02:06  \\
12 &   282 & 00:48 & 01:31  & 02:19  \\ 
24 &   141 & 00:48 & 02:11  & 02:59  \\
32 &   106 & 00:48 & 02:27  & 03:15  \\
\hline
    \multicolumn{5}{l}{{$\dagger$} {Duration in (hh:mm) format.} }\\
    \end{tabular}  
    \end{center}
\end{table}

\begin{figure}
    \begin{center}
    \includegraphics[width=0.9\columnwidth]{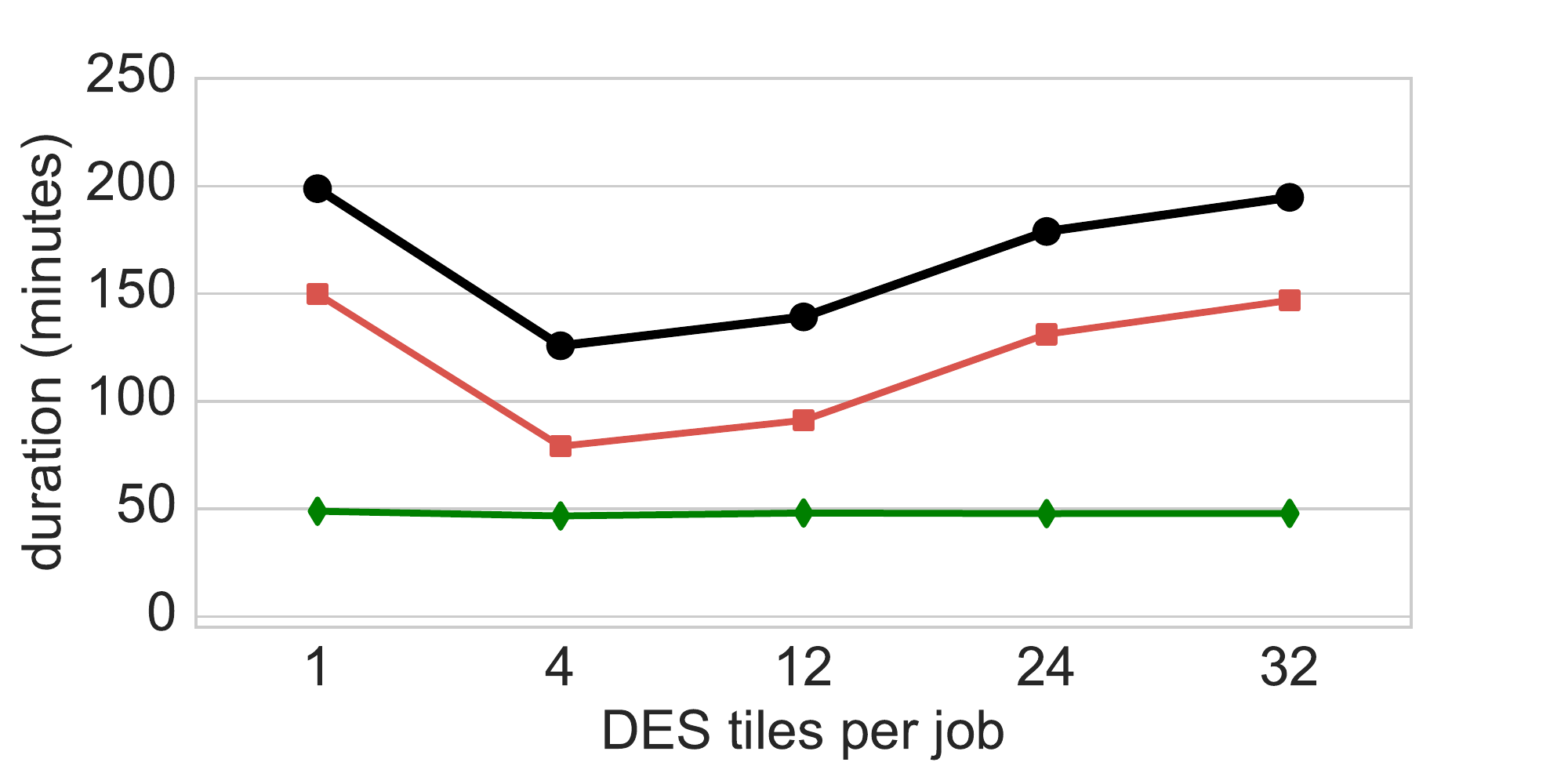}
    \caption{Duration of pipeline \textit{\textbf{Photo-z Compute}} execution as a function of the number of DES tiles processed by each job in parallel. In black circles, the total duration of the runs. In green diamonds, the sum of the durations of all serial components. In red squares, the time spent with the parallelized part of the process, i.e., the actual photo-$z$ estimations by DNF (see Figure~\ref{fig:photoz_flowchart}). }
    \label{fig:pz_compute_by_tiles_SPT}
    \end{center}
\end{figure}

As done for the previous pipelines, we show an example of running \textit{\textbf{Photo-z Compute}} in the supplemental video V4\footnote{\url{https://youtu.be/IcCk0MYhy-E?list=PLGFEWqwqBauBIYa8H6KnZ4d-5ytM59vG2}}.

	    \subsection{Photo-z PDF} 
	    \label{subsec:pz_pdf}
 
The use of redshift probability density functions (PDFs), instead of single estimates of photometric redshifts, is a necessary approach to incorporate the measurement's uncertainties on scientific analyzes. For large astronomical surveys, the storage of billions of PDFs can be a challenge, furthermore if they are measured several times, as when using different methods. 

To overcome these issues, we adopt two procedures when dealing with PDFs in the Portal. The first one is to compute PDFs only for objects selected in a science--ready catalog, avoiding wasting time and storage space with PDFs for stars or bad data. That is the reason for the pipeline not to be present in Figure~\ref{fig:photoz_flowchart}. The second procedure is to apply a method for data compression and store just a reduced list of coefficients representing the PDF, instead of the complete PDF for each galaxy. The details of this method are explained in Rau et al. (2018, in preparation). 

In a similar way to the previous pipelines, we show examples of configurations, such as the photo-$z$ code to be used, and the redshift range for the PDF in Figure~\ref{fig:pz_pdf_config}.

    \begin{figure*}
    \begin{center}
\includegraphics[width=1.7\columnwidth]{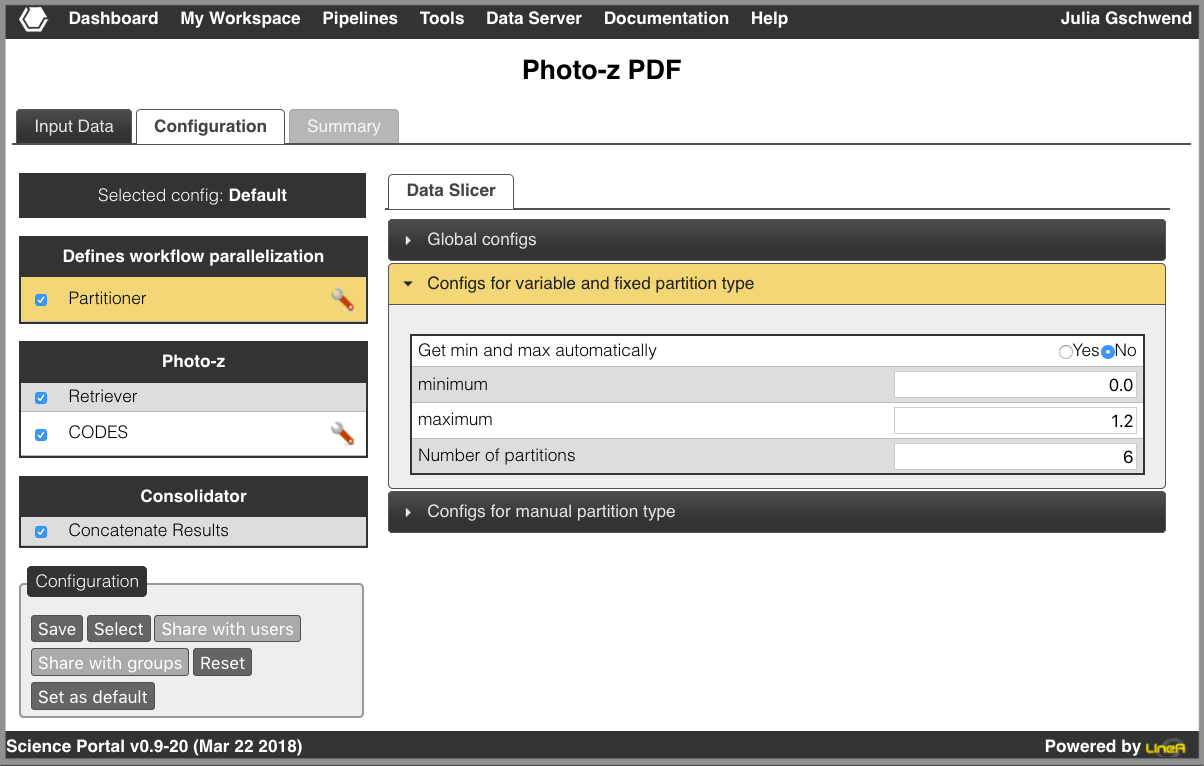}
    \end{center}
    \caption{Screenshot of the configuration tab of the \textit{\textbf{Photo-z  PDF}} pipeline.} 
    \label{fig:pz_pdf_config}
    \end{figure*}

The resulting redshift distribution strongly depends on the stage it is obtained in the Portal, as evident in Figure~\ref{fig:photoz_pdf}, where we show, as an example, for a small sample (dataset COSMOS D04, defined in \ref{app_sub:phot_sample}) obtained using DNF. The pipeline \textit{\textbf{Photo-z Compute}} provides photo-$z$s for every object present in the ``raw'' photometric samples, including stars and poorly sampled objects (left panel). The pipeline responsible for creating science--ready catalogs removes those objects, but its product still contains only point photo-$z$ estimates (middle panel), which can be severely biased. To obtain the final distribution, we do the stacking of the probabilities (right panel) which is considerably smoother than the previous one.

\begin{figure}
\begin{center}
\includegraphics[width=\columnwidth]{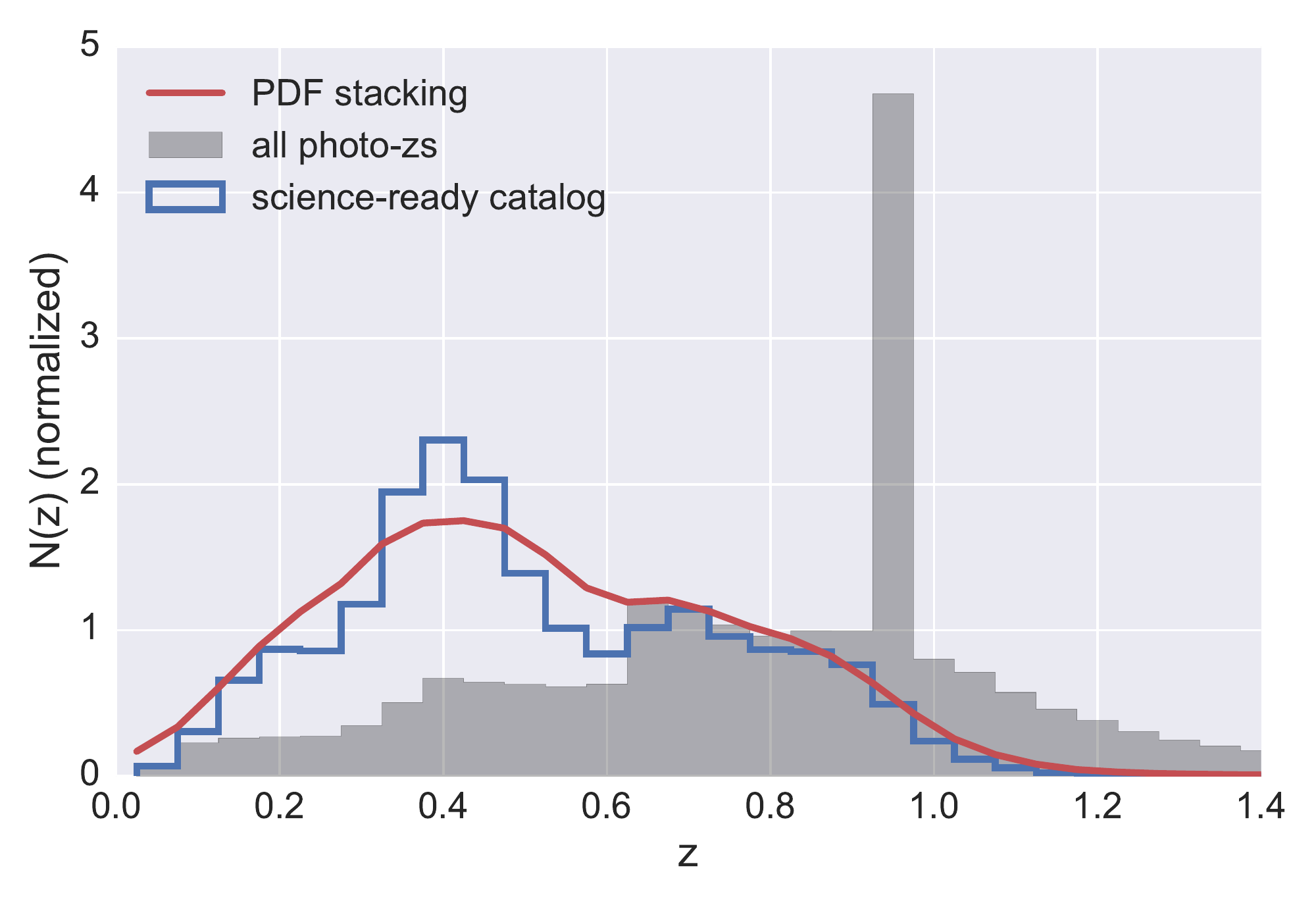} 
\caption{\footnotesize Photo-$z$ distributions of the photometric sample. The stepfilled gray histogram shows the distribution of redshift point estimates for the complete sample (before cleaning from bad data and removing stars). The blue line histogram, still, the point estimates, but after selecting a science--sample. The red line shows the same science sample as the blue line, but considering the probability density functions, $P(z)$, instead of the single estimates.}
\label{fig:photoz_pdf}
\end{center}
\end{figure}

        \section{Summary}
        \label{sec:conclusions}   

In this paper, we describe the infrastructure available in the DES Science Portal to create training sets, training files and to compute photo-$z$ using different algorithms. It is an easy--to--use framework that concatenates different pipelines involved in the calculation of photo-$z$s, ensuring consistency between these processes. 

The database registers all the steps; the Portal framework eases the task of carrying out a large variety of tests and comparing their results. Considering the volume of data, the number of algorithms and the various releases of photometric and spectroscopic data, having a structured framework like the one presented here is critical for vetting of DES algorithmic improvements, and the systematic production of photo-$z$s for future DES releases.

Although the Portal is currently accessible only for the members of DES collaboration, the methodology and lessons learned here can be useful and subject of interest for anyone that uses photo-$z$s in a wide range of science applications.

The database associated with the Portal ingests spectroscopic data regularly. Although the redshift repository continuously grows, the list of surveys used is reported and registered, so that a process can be reproduced or use only the catalogs of interest in another experiment.  

After preparing training sets and photo-$z$s we can compare to spec-$z$s for quality checks. The pipeline used in the estimation of photo-$z$s for large datasets is parallelized to improve performance. The tests presented in Section~\ref{subsec:pz_compute} reveal that a good parallelization strategy is to distribute the data to the CPU cores using data partitions that are small enough to occupy the whole computing cluster, but large enough to avoid creating a queue of idle jobs. The optimal number of tiles or HEALPix pixels to be processed per job is then dependent on the size of the dataset in question and its original data partition. The resulting photo-$z$ tables are amongst the values added in the preparation of catalogs ready for Portal science workflows. 

It is important to point out that the strategy adopted by the Portal is to compute photo-$z$s for all objects in the original catalog produced by DESDM. We do that because the photo-$z$ calculation is, by far, the most computationally intensive step of the E2E process. Calculating photo-$z$s for all objects gives the flexibility to create any catalog for Portal science workflows without having to re--compute photo-$z$s if one decides to change the star/galaxy classifier or another criterion for the sample selection. One disadvantage of our approach is that, in this first pass, we only compute point--values of photo-$z$. 

The calculation of a full PDF happens at a later stage when the number of objects of interest is smaller, after quality pruning and star--galaxy separation. This approach is discussed in a separate paper that focuses on the method of preparing catalogs ready for Portal science workflows \citep{Fau18}. 

For the near future, there will also be pipelines available to be executed through Jupyter Notebooks \citep{Klu16,Per07}, as an alternative to the regular workflow system. There is already a prototype that has been tested using the multi-user web application JupyterHub\footnote{\url https://github.com/jupyterhub}, but the current implementations are not related to photo-$z$s.

All the examples shown in the figures and supplemental videos use data from the Y1A1 data release. Nevertheless, the same infrastructure is valid for any other DES data release and also for simulations. 

Besides already allowing one to handle large datasets and easing a lot of scientific applications, the DES Science Portal has been a useful laboratory of methodologies and a precursor of implementations for the next generation of photometric surveys.

%
    \section*{Acknowledgments}

We thank P. Egeland and F. Ostrovski for the contribution in the early phases of development of this infrastructure. We also thank R. Brito, J.G.S. Dias, V. Machado, L. Nunes, and G. Vila Verde for the contributions in the Portal's basic infrastructure, essential for the realization of this work. 

JG is supported by CAPES. ACR is supported by CNPq process 157684/2015-6. ML is partially supported by CNPq and FAPESP. MA is supported by CNPq process 165049/2017-0. Part of this research is supported by INCT do e--Universo (CNPq grants 465376/2014-2). 

Funding for the DES Projects has been provided by the U.S. Department of Energy, the U.S. National Science Foundation, the Ministry of Science and Education of Spain, the Science and Technology Facilities Council of the United Kingdom, the Higher Education Funding Council for England, the National Center for Supercomputing Applications at the University of Illinois at Urbana-Champaign, the Kavli Institute of Cosmological Physics at the University of Chi\-cago, 
the Center for Cosmology and Astro-Particle Physics at the Ohio State University, the Mitchell Institute for Fundamental Physics and Astronomy at Texas A\&M University, Financiadora de Estudos e Projetos, Funda{\c c}{\~a}o Carlos Chagas Filho de Amparo {\`a} Pesquisa do Estado do Rio de Janeiro, Conselho Nacional de Desenvolvimento Cient{\'i}fico e Tecnol{\'o}gico and the Minist{\'e}rio da Ci{\^e}ncia, Tecnologia e Inova{\c c}{\~a}o, the Deutsche Forschungsgemeinschaft and the Collaborating Institutions in the Dark Energy Survey. 

The Collaborating Institutions are Argonne National Laboratory, the University of California at Santa Cruz, the University of Cambridge, Centro de Investigaciones Energ{\'e}ticas, Medioambientales y Tecnol{\'o}gicas-Madrid, the University of Chi\-cago, University College London, the DES-Brazil Consortium, the University of Edinburgh, the Eidgen{\"o}ssische Technische Hoch\-schule (ETH) Z{\"u}rich, Fermi National Accelerator Laboratory, the University of Illinois at Urbana-Champaign, the Institut de Ci{\`e}ncies de l'Espai (IEEC/CSIC), the Institut de F{\'i}sica d'Altes Energies, Lawrence Berkeley National Laboratory, the Ludwig-Maximilians Universit{\"a}t M{\"u}nchen and the associated Excellence Cluster Universe, the University of Michigan, the National Optical Astronomy Observatory, the University of Nottingham, The Ohio State University, the University of Pennsylvania, the University of Portsmouth, SLAC National Accelerator Laboratory, Stanford University, the University of Sussex, Texas A\&M University, and the OzDES Membership Consortium.

Based in part on observations at Cerro Tololo Inter-American Observatory, National Optical Astronomy Observatory, which is operated by the Association of 
Universities for Research in Astronomy (AURA) under a cooperative agreement with the National Science Foundation.

The DES data management system is supported by the National Science Foundation under Grant Numbers AST-1138766 and AST-1536171. The DES participants from Spanish institutions are partially supported by MINECO under grants AYA2015\-71825, ESP2015-66861, FPA2015-68048, SEV-2016-0588, SEV-2016-0597, and MDM-2015-0509, some of which include ERDF funds from the European Union. IFAE is partially funded by the CERCA program of the Generalitat de Catalunya. Research leading to these results has received funding from the European Research
Council under the European Union's Seventh Framework Program (FP7/2007-2013) including ERC grant agreements 240672, 291329, and 306478. We  acknowledge support from the Australian Research Council Centre of Excellence for All-sky Astrophysics (CAASTRO), through project number CE110001020, and the Brazilian Instituto Nacional de Ci\^encia  e Tecnologia (INCT) e-Universe (CNPq grant 465376/2014-2).

This manuscript has been authored by Fermi Research Alliance, LLC under Contract No. DE-AC02-07CH11359 with the U.S. Department of Energy, Office of Science, Office of High Energy Physics. The United States Government retains and the publisher, by accepting the article for publication, acknowledges that the United States Government retains a non-exclusive, paid-up, irrevocable, world-wide license to publish or reproduce the published form of this manuscript, or allow others to do so, for United States Government purposes.

\bibliography{main}

\appendix

    \section{Data description}
    \label{app_sec:data}

As a proof of concept, we show through this paper an example of the sequence of pipelines run to estimate photo-$z$ and use these results to discuss the benefits of such infrastructure. The results presented in Section~\ref{sec:pz_pipe}, after each pipeline methodology was described. In the following sections, we briefly describe the data used in those runs.

    \subsection{Photometric data}
    \label{app_sub:phot_sample}

To describe the processes carried out in the Portal to estimate photo-$z$s, we use photometric data from the first annual internal release of DES. The observations were carried out with the mosaic camera DECam \citep{Fla15,Hon14} built as part of DES project and mounted on the 4-meter Blanco telescope at Cerro Tololo Inter-American Observatory (CTIO), in Chile. 

The data were reduced and calibrated by the DES Data Management (DESDM) team at the National Center for Supercomputing Applications (NCSA) using standard procedures descri\-bed by \citet{Des12}, \citet{Moh12}, \citet{Mor18}. This is the system used for the processing and calibration of DES data, and the DECam Community Pipeline. The observations \citep{Die14} reported here took place from August 2013 to February 2014 and include a total of 14,340 exposures in the $grizY$ filters, covering a total area of $\sim$1,800 deg$^2$ in eight distinct regions, making the so-called DES Y1A1 release \citep{Drl18}.

The two largest regions (see Figure~\ref{fig:location-allfields}) are part of the wide-field survey. One of about 160 deg$^2$ overlapping the Sloan Digital Sky Survey Stripe 82 Imaging Data \citep[S82,][]{Jia2014}, and another of $\sim$1,400 deg$^2$ overlapping the region observed by the South Pole Telescope \citep[SPT,][]{Car11}. These two wide regions were covered with up to four passes in each filter, reaching SExtractor's \texttt{mag\_auto} magnitude limits of $g=23.4$, $r=23.2$, $i=22.5$, $z=21.8$, and $Y=20.1$ \citep{Drl18} in the AB system for a 10$\sigma$ detection limit. 

    \begin{figure*}
    \begin{center}
   \includegraphics[width=1.4\columnwidth]{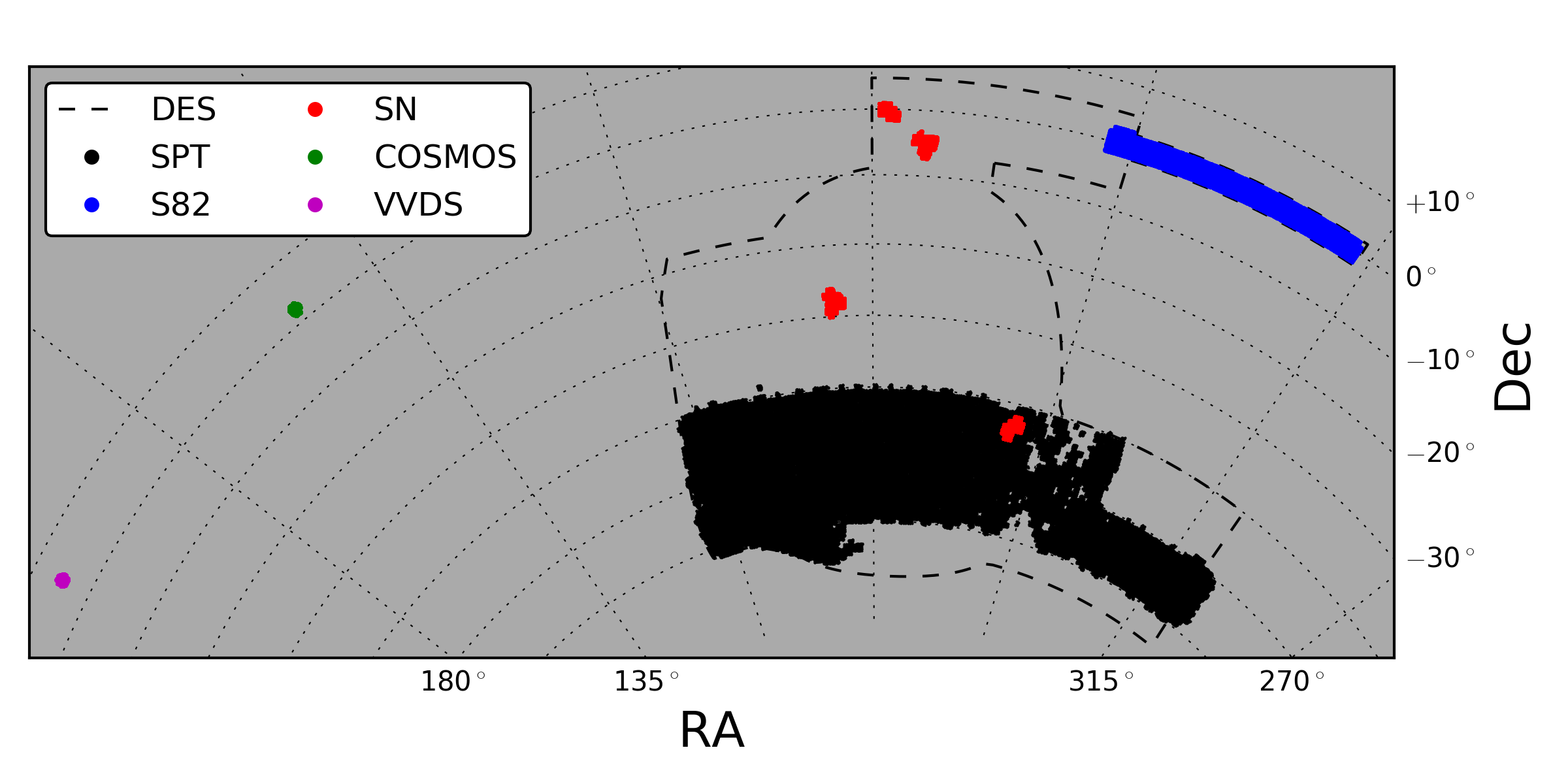}
    \end{center}
    \caption{Location of all the Y1A1 fields used in this paper and DES footprint (dashed line).} 
    \label{fig:location-allfields}
    \end{figure*}

The remaining regions, called ``supplemental fields'' -- where a large number of spectroscopic redshifts (spec-$z$s) are available -- belong to both the science verification phase\footnote{\url{https://des.ncsa.illinois.edu/releases/sva1}} (SVA1), and Y1A1 releases. Four of these regions are collectively known as Supernova (SN) fields. One of the other regions overlaps with the VVDS-14h field from VIMOS VLT Deep Survey \citep[hereafter VVDS,][]{LeF05} and the final region overlaps with COSMOS field \citep{Sco07}. The SN fields are regularly observed as part of the SNe Ia program, making available a greater number of exposures compared to the wide survey. The location of all these regions are shown in Figure~\ref{fig:location-allfields} along with the DES footprint. Relevant information is available in Table~\ref{table:des_fields}.

\begin{table*}
    \begin{center}
    \caption{Basic information of DES Y1 photometric datasets.} 
    \label{table:des_fields}
    \begin{tabular}{l|rrr}
\hline
\ \ \  Field  & \# Objects\ \ \ & 
Area\textsuperscript{$\dagger$} \  & mag lim\textsuperscript{$\ddagger$}   \\
\hline
COSMOS &     313,380 &      2.97 & 23.6 \\   
SN     &   2,569,018 &     31.76 & 24.4 \\ 
SPT    & 126,623,762 &  1,469.05 & 23.2 \\ 
VVDS   &     260,446 &      2.91 & 23.6 \\  
S82    &  12,487,566 &    165.84 & 23.4 \\   
\hline
    \multicolumn{4}{l}{{$\dagger$}\footnotesize{\ Covered area in deg$^2$}} \\
    \multicolumn{4}{l}{{$\ddagger$}\footnotesize{\ Defined as the peak of $i$-band \textsc{mag\_auto} number counts}} \\
    \end{tabular}  
    \end{center}
\end{table*}

    \subsection{Spectroscopic data}
    \label{app_sub:spec_data}
 
In this paper, we use a spec-$z$ sample with reliable measurements to train photo-$z$ algorithms, and to test their performance, as an example of validation procedure. The construction of this sample is by compiling data available from a large number of surveys individually ingested into the database associated with DES Science Portal. 

Currently, the Portal's spectroscopic database contains redshift measurements from a total of 34 spectroscopic surveys. Together, these catalogs contain 2,173,561 redshift measurements, where 1,688,403 refers to extragalactic sources including both galaxies and quasars. In Table~\ref{table:all_specs} we show the information about the spectroscopic sample used as an example in Section~\ref{subsec:spec_sample}. The surveys ordered by the number of successful matchings with photometric data from DES Y1 wide fields are the numbers in the fourth column. The second column shows the number of objects with good quality spec-$z$s per survey, after resolving multiple measurements as discussed in Section~\ref{subsec:spec_sample}. 

After dealing with quality cuts and multiple measurements in the spectroscopic database, we end up with 1,412,816 unique high quality spec-$z$s. However, we know that not all of these sources will be matched to the photometric sample since they extend beyond the Y1A1 DES footprint. In particular, around 170 thousand sources overlap with Y1 footprint. 

    \begin{table*}
    \begin{center}
    \begin{threeparttable}[b]
    \centering
    \caption{Spectroscopic samples used in this paper.}  
	\footnotesize
    \label{table:all_specs}
    \begin{tabular}
    {lrrcrcccc}
\hline                    
Survey \    & \ \  \# objects\textsuperscript{$\dagger$}  &\ \  \%  \ \  & \# matchings &  \ \  \% \ \ &    $z$ mean     &    $z$ min   &  $z$ max  & \ \ \  Ref. \textsuperscript{$\ddagger$}\\
\hline 
PRIMUS      & 110,522 &  7.8 & 60,477 & 35.3 & 0.57 & 0.02 & 4.08 & 1   \\
SDSS DR14   & 423,353 & 30.1 & 19,778 & 11.5 & 0.59 & 0.00 & 1.95 & 2   \\
DES AAOmega & 23,114  &  1.6 & 16,492 &  9.6 & 0.54 & 0.00 & 3.94 & 3   \\
VIPERS      & 48,558  &  3.4 & 14,832 &  8.6 & 0.69 & 0.05 & 1.67 & 4   \\
WiggleZ     & 80,431  &  5.7 & 9,131  &  5.3 & 0.55 & 0.01 & 1.70 & 5   \\
VVDS        & 13,638  &  1.0 & 7,532  &  4.4 & 0.59 & 0.00 & 4.08 & 6   \\
zCOSMOS     & 12,513  &  0.9 & 7,511  &  4.4 & 0.54 & 0.00 & 1.99 & 7   \\
3D-HST      & 180,841 & 12.8 & 6,333  &  3.7 & 1.01 & 0.01 & 5.21 & 8   \\
DEEP2       & 33,936  &  2.4 & 5,402  &  3.1 & 0.99 & 0.01 & 1.89 & 9   \\
2dF         & 211,705 & 15.0 & 3,547  &  2.1 & 0.12 & 0.00 & 0.35 & 10  \\
GAMA        & 7,429   &  0.5 & 3,444  &  2.0 & 0.22 & 0.01 & 0.74 & 11  \\
ACES        & 4,047   &  0.3 & 3,045  &  1.8 & 0.58 & 0.04 & 1.42 & 12  \\
6dF         & 108,760 &  7.7 & 2,637  &  1.5 & 0.06 & 0.00 & 0.38 & 13  \\
DES IMACS   & 2,387   &  0.2 & 2,215  &  1.3 & 0.60 & 0.00 & 1.37 & 14  \\
SAGA        & 64,033  &  4.5 & 1,994  &  1.2 & 0.29 & 0.01 & 1.17 & 15  \\
NOAO OzDES  & 3,008   &  0.2 & 1,884  &  1.1 & 0.22 & 0.00 & 0.68 & 16  \\
XXL AAOmega & 3,143   &  0.2 & 926   &  0.5 & 0.47 & 0.00 & 2.80 & 17  \\
SPT GMOS    & 2,189   &  0.2 & 790   &  0.5 & 0.56 & 0.07 & 1.24 & 18  \\
UDS         & 1,307   &  0.1 & 705   &  0.4 & 1.06 & 0.04 & 3.44 & 19  \\
SNLS FORS   & 1,321   &  0.1 & 529   &  0.3 & 0.51 & 0.03 & 3.75 & 20  \\
ATLAS       & 729     &  0.1 & 503   &  0.3 & 0.32 & 0.02 & 1.89 & 21  \\
Pan-STARRS  & 1,775   &  0.1 & 463   &  0.3 & 0.33 & 0.00 & 3.16 & 22  \\
C3R2        & 1,249   &  0.1 & 429   &  0.3 & 0.92 & 0.03 & 3.52 & 23  \\
SpARCS      & 403     &  $<$0.1 & 356   &  0.2 & 0.91 & 0.12 & 1.58 & 24  \\
SNVETO      & 2,154   &  0.2 & 178   &  0.1 & 0.84 & 0.03 & 3.63 & 25  \\
FMOS-COSMOS & 328     &  $<$0.1 & 173   &  0.1 & 1.55 & 0.75 & 1.74 & 26  \\
SNLS AAOmega& 350     &  $<$0.1 & 58  &  $<$0.1 & 0.60 & 0.07 & 1.17 & 27  \\
CDB         & 388    &  $<$0.1 & 38    &  $<$0.1 & 0.58 & 0.25 & 0.91 & 28  \\
VUDS        & 141    &  $<$0.1 & 36    &  $<$0.1 & 1.78 & 0.19 & 3.75 & 29  \\
ZFIRE       & 202    &  $<$0.1 & 29    &  $<$0.1 & 1.77 & 1.05 & 2.26 & 30  \\
MOSFIRE     & 143    &  $<$0.1 & 25    &  $<$0.1 & 1.89 & 0.80 & 3.08 & 31  \\
2dFLenS     & 63,632  &  4.5 & 23   &  $<$0.1 & 0.40 & 0.09 & 0.69 & 32  \\
GLASS       & 383    &  $<$0.1 & 10    &  $<$0.1 & 1.06 & 0.34 & 2.07 & 33  \\
XMM-LSS     & 26     &  $<$0.1 & 5     &  $<$0.1 & 0.42 & 0.19 & 0.65 & 34  \\
\hline
    \end{tabular}
    \begin{tablenotes}
	\scriptsize
   \item $\dagger$ Only selected objects with $Q_{spec}\geqslant 3$
\item  $\ddagger$ References: 
1- \citet{Coi11,Coo13} and \url{https://primus.ucsd.edu/}; 
2- \citet{Abo17arXiv} and \url{http://www.sdss.org/dr14/};
3- \citet{Yua15,Chi17}; 
4- \citet{Gar14} and \url{http://vipers.inaf.it/rel-pdr1.html}; 
5- \citet{Par12} and \url{http://wigglez.swin.edu.au/site/};
6- \citet{Gar08,LeF04};
7- \citet{Lil09}; 
8- \citet{Mom16} and \url{http://3dhst.research.yale.edu/Data.php}; 
9- \citet{Dav03,Dav07} and  \url{http://deep.ps.uci.edu/DR4/home.html}; 
10- \citet{Col01}  \url{http://www.2dfgrs.net/};  
11- \citet{Dri11};  
12- \citet{Coo12} and \url{http://mur.ps.uci.edu/cooper/ACES/zcatalog.html}; 
13- \citet{Jon09} and \url{http://www.6dfgs.net/}; 
14- \citet{Nor16};  
15- \citet{Geh17} and \url{http://sagasurvey.org/}; 
16- \citet{Yua15,Chi17};
17- \citet{Lid16} and \url{http://cosmosdb.iasf-milano.inaf.it/XXL/}; 
18- \citet{Bay16} and  \url{https://dataverse.harvard.edu/dataset.xhtml?persistentId=doi:10.7910/DVN/OR13NN/};
19- \url{http://www.nottingham.ac.uk/astronomy/UDS/UDSz/};  
20- \citet{Baz11} Private communication; 
21- \citet{Mao12};  
22- \citet{Res14,Sco14,Kai10};  
23- \citet{Mas17};
24- \citet{Muz12}; 
25- \url{http://www.ast.cam.ac.uk/~fo250/Research/SNveto/};
26- \citet{Sil15} and \url{http://member.ipmu.jp/fmos-cosmos/FC\_catalogs.html};
27- \citet{Lid13,Yua15,Chi17}  and \url{http://apm5.ast.cam.ac.uk/arc-bin/wdb/aat\_database/observation\_log/make};  
28- \citet{Sul11};
29- \citet{Tas17} and \url{http://cesam.lam.fr/vuds/DR1/}; 
30- \citet{Nan16} and \url{http://zfire.swinburne.edu.au/data.html};
31- \url{http://mosdef.astro.berkeley.edu};
32- \citet{Bla16} and \url{http://2dflens.swin.edu.au/};
33- \citet{Tre15} and \url{https://archive.stsci.edu/prepds/glass/};
34- \citet{Sta10}. 
    \end{tablenotes}
    \end{threeparttable}
    \end{center}
    \end{table*}
%


\end{document}